\documentclass[a4paper,fleqn,usenatbib]{mnras}

\usepackage{newtxtext,newtxmath}

\usepackage[T1]{fontenc}
\usepackage{ae,aecompl}

\usepackage[dvipdfmx]{graphicx}	
\usepackage{amsmath}	
\usepackage{indentfirst}	
\usepackage{pict2e}
\usepackage{color}
\usepackage{bm}
\usepackage{tablefootnote}
\usepackage{pict2e}

\usepackage{ulem}
\definecolor{ao_en}{rgb}{0.0, 0.5, 0.0}

\defcitealias{2020MNRAS.496.2762F}{Paper I}

\title[Generation of circular polarization of Lyman $\alpha$]{
Generation of high circular polarization of interstellar Lyman $\alpha$ radiation triggering biological homochirality   
}

\author[Fukushima et al.]{
Hajime Fukushima$^{1}$\thanks{E-mail:fukushima@ccs.tsukuba.ac.jp},
Hidenobu Yajima$^{1}$,
Masayuki Umemura$^{1}$
\\
$^{1}$Center for Computational Sciences, University of Tsukuba, Ten-nodai, 1-1-1 Tsukuba, Ibaraki 305-8577, Japan\\
}
\date{Accepted XXX. Received YYY; in original form ZZZ}

\pubyear{2019}

\begin{document}
\label{firstpage}
\pagerange{\pageref{firstpage}--\pageref{lastpage}}
\maketitle

\begin{abstract}
The homochirality of biological molecules on the Earth is a long-standing mystery regarding the origin of life. 
Circularly polarized ultraviolet (UV) light could induce the enantiomeric excess of biological molecules in the interstellar medium, leading to the homochirality on the earth.
By performing 3D radiation transfer simulations with multiple scattering processes in interstellar dusty slabs, we study the generation of circular polarization (CP) of ultraviolet light at Lyman $\alpha$ ($\lambda = 0.1216~{\rm \mu m}$) as well as in the near-infrared (NIR, $\lambda = 2.14~{\rm \mu m}$) wavelengths.
Our simulations show that the distributions of CP exhibit a symmetric quadrupole pattern, regardless of wavelength and viewing angle.
The CP degree of scattered light from a dusty slab composed of aligned grains is $\sim 15$ percent for Ly$\alpha$ and $\sim 3$ percent at NIR wavelengths in the case of oblate grains with an MRN size distribution.
We find that the CP degree of Ly$\alpha$ is well correlated with that in the NIR regardless of viewing angles, whilst being a factor of $\sim 5$ higher.
Thus, high CP of Ly$\alpha$ is expected in sites where NIR CP is detected. 
We suggest that such circularly polarized Ly$\alpha$ may initiate the enantiomeric excess of biological molecules in space.

\end{abstract}



\section{Introduction}\label{introduction} 

The homochirality of biological molecules on the Earth is a long-standing unrevealed issue for the origin of life. 
Biological molecules including amino acids are detected in meteorites, such as the Murchison and Mukundpura meteorites \citep{1997Natur.389..265E, 2018P&SS..164..127P}.
Also, the Hayabusa2 spacecraft recently collected the samples of Ryugu \citep{yokoyama2022samples}, and amino acids were discovered \citep{nakamura2022origin}. Besides, the enantiomeric excess of the L-amino acids is found in the Murchison and Mukundpura meteorites.
It is proposed that circularly polarized light can play a crucial role in the enantiomeric excesses of biological molecules \citep[see][for a review]{1991OLEB...21...59B}. Thereafter,
\citet{1998Sci...281..672B} demonstrated that the circularly polarized ultraviolet (UV) light in the space can induce the enantiomeric excesses of biological molecules discovered on the meteorites.

For now, the observations of circular polarization (CP) caused by dust grains have been limited at near-infrared (NIR) wavelengths and various degrees of CP have been detected around young stellar objects
\citep[e.g.,][]{2000MNRAS.312..103C, 2010OLEB...40..335F, 2014ApJ...795L..16K, 2016AJ....152...67K, 2018AJ....156....1K}.
The NIR CP degree is correlated with the luminosity of the central stars, and high CP of more than 10 percent has only been observed around massive stars \citep{2014ApJ...795L..16K}.
The previous works showed that aspherical dust grains efficiently produce the CP light \citep{2000MNRAS.314..123G, 2002A&A...385..365W, 2002ApJ...574..205W, 2004MNRAS.352.1347L}.
In \citet[][hearafter Paper I]{2020MNRAS.496.2762F}, we have successfully reproduced the quadrupole patterns of the observed CP maps at NIR wavelengths and suggested that the CP degree larger than 10 percent could be generated by micron-sized dust grains.

On the other hand, \citet{2005OLEB...35...29L} theoretically investigated the UV circular polarization of scattered light around a massive star.
They concluded that the CP degree of singly scattered light is less than 10 percent, and that dichroic extinction with multiple scatterings is needed to produce higher CP.
However, they only considered the case with the wavelength range larger than $\lambda = 0.2~{\rm \mu m}$.
The CP degree at the shorter wavelength is still unclear.

The absorption of CP light by chiral molecules is caused
by the circular dichroism (CD).
If a continuous flux exists in a wide range 
of wavelengths, the CD alternates in sign and sums to zero
over the whole spectrum
due to the Kuhn-Condon sum rule \citep[][]{kuhn1930physical,RevModPhys.9.432}.
Therefore, to initiate the enantiomeric excess, a circularly polarized strong emission line at a specific wavelength is requisite.
A possibility is Ly$\alpha$ lines from hydrogen
since a lot of distant galaxies with strong Ly$\alpha$ line emissions from hydrogen have been observed \citep{2020ARA&A..58..617O}.
The solar system, which is located at the outskirt of the Milky Way,
could be exposed to strong Ly$\alpha$ radiation from central star-forming regions in early galactic phase.
Also, high-redshift Ly$\alpha$ emitting galaxies
may produce strong Ly$\alpha$ background radiation
in addition to direct radiation from star-forming regions.
Ly$\alpha$ photons ($\lambda= 0.1216~{\rm \mu m}$) are produced via the hydrogen recombination in ionized gas
\citep{2009MNRAS.398..715Y, 2014MNRAS.440..776Y}. 
In the star cluster formation, photoionization feedback from massive stars is the main mechanism to disrupt star-forming clouds \citep[e.g.,][]{2020MNRAS.497.3830F, 2021MNRAS.506.5512F, 2022MNRAS.511.3346F}.
Most stars are born as members of a star cluster \citep{2003ARA&A..41...57L}, and sun-like stars are also surrounded by ionized gas at birth.
Therefore, the early solar system could form under the environment with strong Ly$\alpha$ radiation. 
Recently, it is suggested that circularly polarized Ly$\alpha$ is favorable to cause the enantiomeric excess of the precursors and the amino acids \citep{shoji2023enantiomeric} 
and propylene oxide \citep{hori2022theoretical} detected in the Sagittarius B2 star-forming region \citep{2016Sci...352.1449M}.
However, the generation of Ly$\alpha$ CP has not been elucidated so far.

In this study, we explore the generation of CP at Lyman $\alpha$ ($\lambda = 0.1216~{\rm \mu m}$).
We utilize the 3-dimensional radiation transfer code developed in \citetalias{2020MNRAS.496.2762F}.
The simulations follow changes of the Stokes parameters at each scattering and absorption process.
We also investigate the dependence of the CP distributions on the observational angles.
In Section \ref{Sec:numerical_method}, we describe the numerical method and setups of our simulations.
The main results are presented in Section \ref{Sec:results}.
Section \ref{sec:summary_discussion} is devoted to summary and discussion.

\section{NUMERICAL METHOD}\label{Sec:numerical_method}

We utilize the radiation transfer code based on the Monte Carlo technique developed in \citetalias{2020MNRAS.496.2762F} that follows a change of the Stokes parameters. 
We consider the UV wavelength ($\lambda = 0.1216~{\rm \mu m}$), which corresponds to Ly$\alpha$ photons.
We calculate scattering and absorption processes in a dusty slab around a single radiation source.

In the following subsections, we describe the brief summary of radiation transfer methods in Section \ref{radiation_transfer}, the models of dust grains in Section \ref{Dust_grains}, and the setup of the dusty slab in Section \ref{dusty_slab}.
In Appendix \ref{appendix_scattering}, we describe details of the simulation method. 
We also show the results at the NIR wavelength ($\lambda = 2.14~{\rm \mu m}$) in Appendix \ref{sec:large_dust_grain_NIR}
to compare CP at Ly$\alpha$.

\subsection{Radiation transfer}\label{radiation_transfer}

Photons emitted from stars propagate with absorption and scattering due to dust grains that change their polarization properties. In our simulations,
each photon packet contains the information of the stokes vectors ${\bf{I}} = (I, Q, U, V)^{T}$.
We conduct the Monte Carlo radiation transfer simulations in the following steps:
\begin{description}
	\item[{\it Photon injection}:] 
    In this study, we assume that radiation from a star is unpolarized initially, i.e., the Stokes vector is ${\bf I} = (I_{0}, 0, 0, 0)$ where $I_0$ is the intensity at the surface of radiation sources.
    The initial direction of a photon packet is chosen randomly.
    Then, the radiation transfer calculation starts.
	 
	\item[{\it Step 1}:] 
	We set the optical depth $\tau_{\rm sc, p}$ between the current position and the next scattering.
    The optical depth is evaluated as $\tau_{\rm sc, p} = - \ln(R)$, where $R$ is the random number in the range $[ 0 - 1 ]$.
	
	\item[{\it Step 2}:] 
	We calculate the photon propagation until the total optical depth $\tau_{\rm sc}$ is equal to  $\tau_{\rm sc, p}$.
	In each cell, we estimate the scattering cross-section of dust grains and add it to the total optical depth $\tau_{\rm sc}$ by considering the number density of dust grains and the path length over the cell.
    If $\tau_{\rm sc}$ is still smaller than $\tau_{\rm sc, p}$, we consider the dust extinction in the cell as Equation (3) in \citetalias{2020MNRAS.496.2762F}.
    
	\item[{\it Step 3}:] 
    Once $\tau_{\rm sc}$ reaches $\tau_{\rm sc, p}$, we derive the changes of the Stokes vector and the propagation angle.
    The number of angle bins is 1024. The scattering angle is randomly selected with the weight of the differential cross-section $d C_{\rm sca}/ d \Omega$.
	Depending on the scattering angle, the Stokes vector is determined.
    Finally, we reset $\tau_{\rm sc}$ and $\tau_{\rm sc, p}$, and we return to {\it Step 1}. 
    We consider the radiative transfer to a mock observer at each scattering process for constructing a two-dimensional image as Equation (4) in \citetalias{2020MNRAS.496.2762F}.
	
	\item[{\it Final step}:] 
    When a photon packet reaches the boundary of the computational box, we stop the radiative transfer. Then we start to calculate the radiative transfer of the next photon packet. 
	 
\end{description}

In {\it Step 3}, the Stokes vector after a scattering process is defined with the M\"uller matrix as \citep[e.g.,][]{1983asls.book.....B},
\begin{eqnarray}
\left( \begin{array}{c} 
I\\
Q \\ 
U \\
V
\end{array} \right)_{\rm sc} =  \frac{1}{k^2r^2} \left(
\begin{matrix} 
S_{11}  &  S_{\rm 12} & S_{\rm 13} & S_{\rm 14} \\
S_{21}  &  S_{\rm 22}& S_{23} & S_{24} \\ 
S_{31}  &  S_{\rm 32}& S_{33} & S_{34} \\ 
S_{41}  &  S_{\rm 42}& S_{43} & S_{44}
\end{matrix}
\right)
\left( \begin{array}{c} 
I \\
Q \\ 
U \\
V
\end{array} \right)_{\rm i}, \label{1.1}
\end{eqnarray}
where $k$ and $r$ are the wave number of photons and the distance from the scattering point.
The M\"uller matrix depends on the photon propagation and scattering angles as well as the opening angle between the incident light and the minor axis of dust grains. 
As discussed in Section \ref{Dust_grains}, we assume that the shape of dust grains is oblate, that is rotationally symmetric around the minor axis.
More details of {\it Step 3} is described in Appendix \ref{appendix_scattering}.

In this work, we conduct the simulations with $10^8$ photon packets. 
The propagation angle of photons packets is discretized by \textsc{HEALPIX} \citep{2005ApJ...622..759G} with 1024 angle bins.

\subsection{Dust grain model}\label{Dust_grains}
We adopt a dust model with oblate spheroidal grains with axis ratio of 2:1 as in \citetalias{2020MNRAS.496.2762F}.
We assume the size distribution of the MRN dust mixture 
\citep{1977ApJ...217..425M} in this study.
The number density of dust grains $(n_{\rm d})$ and the range of  dust radii $(a_{\rm d})$ are given as $n_{\rm d} \propto a_{\rm d}^{-3.5}$ and $0.005~{\rm \mu m} \leqq a_{\rm d} \leqq 0.25~{\rm \mu m}$.
Here, the radius of a dust grain is defined as $a_{\rm d} = (3V_{\rm d}/4 \pi )^{1/3}$ where $V_{\rm d}$ is the volume.
In \citetalias{2020MNRAS.496.2762F}, we found that the micron-size dust grains mainly produce highly circularly polarized light at the NIR wavelength.
In Appendix \ref{sec:large_dust_grain_NIR}, 
for viewing angles other than that adopted in \citetalias{2020MNRAS.496.2762F},
we show the results at the NIR wavelength with the MRN dust mixture and the micron-size dust grains.

For the radiative transfer calculations, we prepare tables of the M\"uller matrix (Eq. \ref{1.1}) and the absorption rates (Eq.3 in \citetalias{2020MNRAS.496.2762F}) in advance. 
We use {\sc DDSCAT} code \citep{2013arXiv1305.6497D} in which a dust grain is treated as an array of electric dipoles.
The code allows us to calculate the optical properties at each scattering angle of ($\theta_{\rm s}, \phi_{\rm s}$), and the angle between the minor axis of a dust grain and the direction of incident light ($\Theta$).
The scattering angles are discretized with 1024 angle bins.
The optical properties of dust grains depend not only on the shape but also on the dielectric function.
We adopt the values of dielectric function of astronomical silicate in \citet{2003ApJ...598.1017D}, $m=1.9+0.7i$ for $0.1216~{\rm \mu m}$ and $m=1.7+0.034 i$ for $\lambda = 2.14 ~{\rm \mu m}$.

\subsection{Setup of dusty slab}\label{dusty_slab}
\begin{figure}
	\begin{center}
	\includegraphics[width=\linewidth]{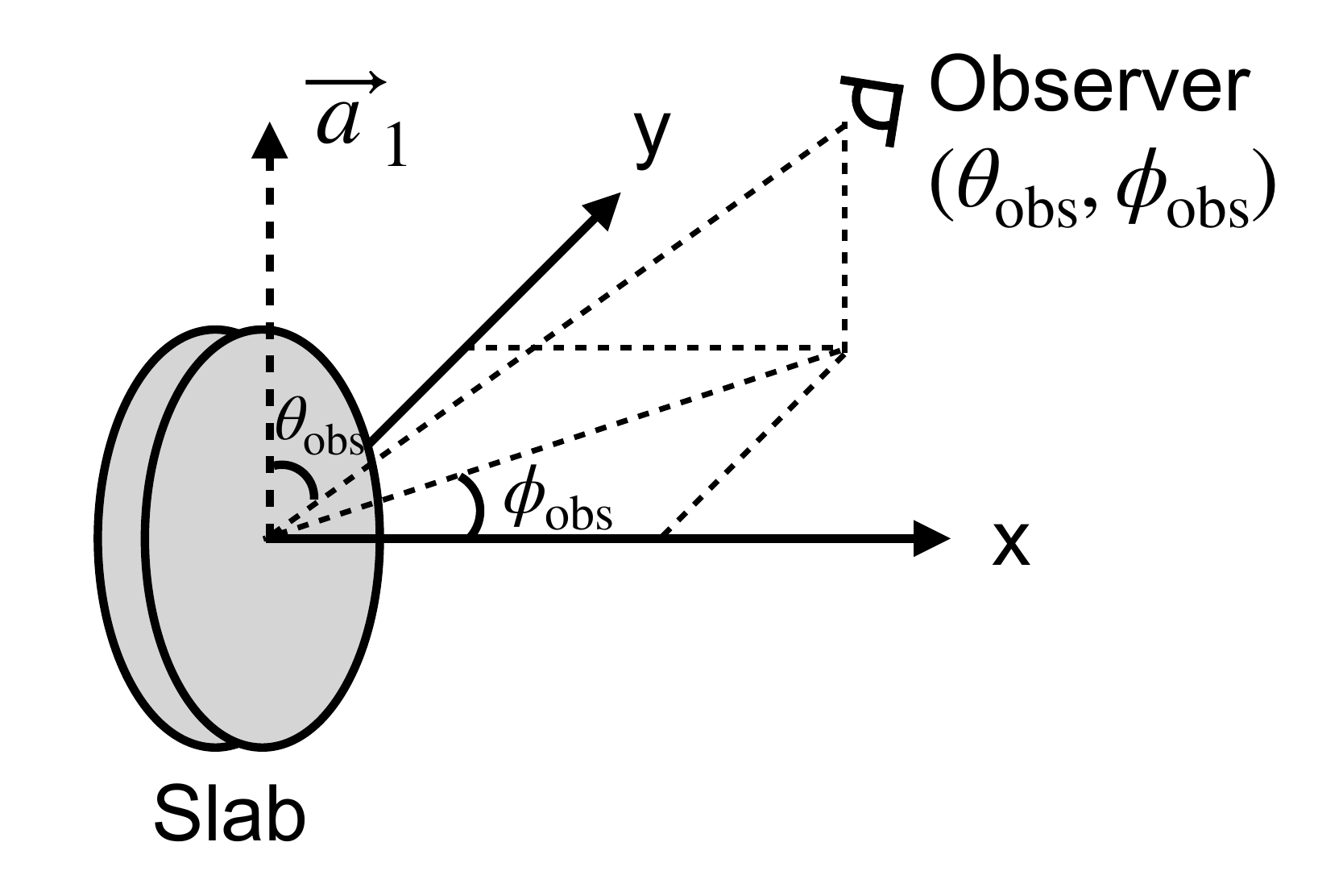}
	\end{center}
	\caption{
	The spatial positions of the dusty slab and the observer.
	The slab is along the yz-plane. 
	The vector $\vec{a}_1$ represents the direction of the minor axis of dust grains.
	The angles $(\theta_{\rm obs}, \phi_{\rm obs})$ are the position of the observer.
	}
	\label{fig_slab}
\end{figure}

We set a dusty slab with a thickness of $0.2~{\rm pc}$ and the radiation source at the centre, as shown in Fig. \ref{fig_slab}.
In this work, we adopt simple geometry to understand the origin of CP clearly. We consider a typical scale of filament induced by turbulence motion as the thickness of the dusty slab \citep[e.g.,][]{2007ARA&A..45..565M, 2011A&A...529L...6A, 2020MNRAS.497.3830F}.
Here, we assume that the minor axis direction of the oblate dust grains ($\vec{a}_{1}$) is aligned along the $z$-axis in the Lab-frame, $\vec{a}_1 = (0,0,1)$.
The number density of dust grains is set as the scattering optical depth along the horizontal direction becomes 1.0.
The slab is resolved by $300 \times 300 \times 60$ cells.
The dust column density with the optical depth is reasonable considering the typical molecular clouds with the surface gas density of $\sim 100 ~\rm M_{\odot}~pc^{-2}$ \citep[e.g.,][]{2010ApJ...723..492R}.

In \citetalias{2020MNRAS.496.2762F}, we only considered the face-on view, which corresponds to the angle $(\theta_{\rm obs}, \phi_{\rm obs})=(90^{\circ}, 0^{\circ})$.
In this study, we consider observational maps of five viewing angles of $(\theta_{\rm obs}, \phi_{\rm obs})=(90^{\circ}, 0^{\circ})$, $(90^{\circ}, 30^{\circ})$, , $(90^{\circ}, 60^{\circ})$$(60^{\circ}, 0^{\circ})$, and $(30^{\circ}, 0^{\circ})$, to investigate dependence of CP degrees on viewing angles.

\begin{figure*}
	\begin{center}
	\includegraphics[width=160mm]{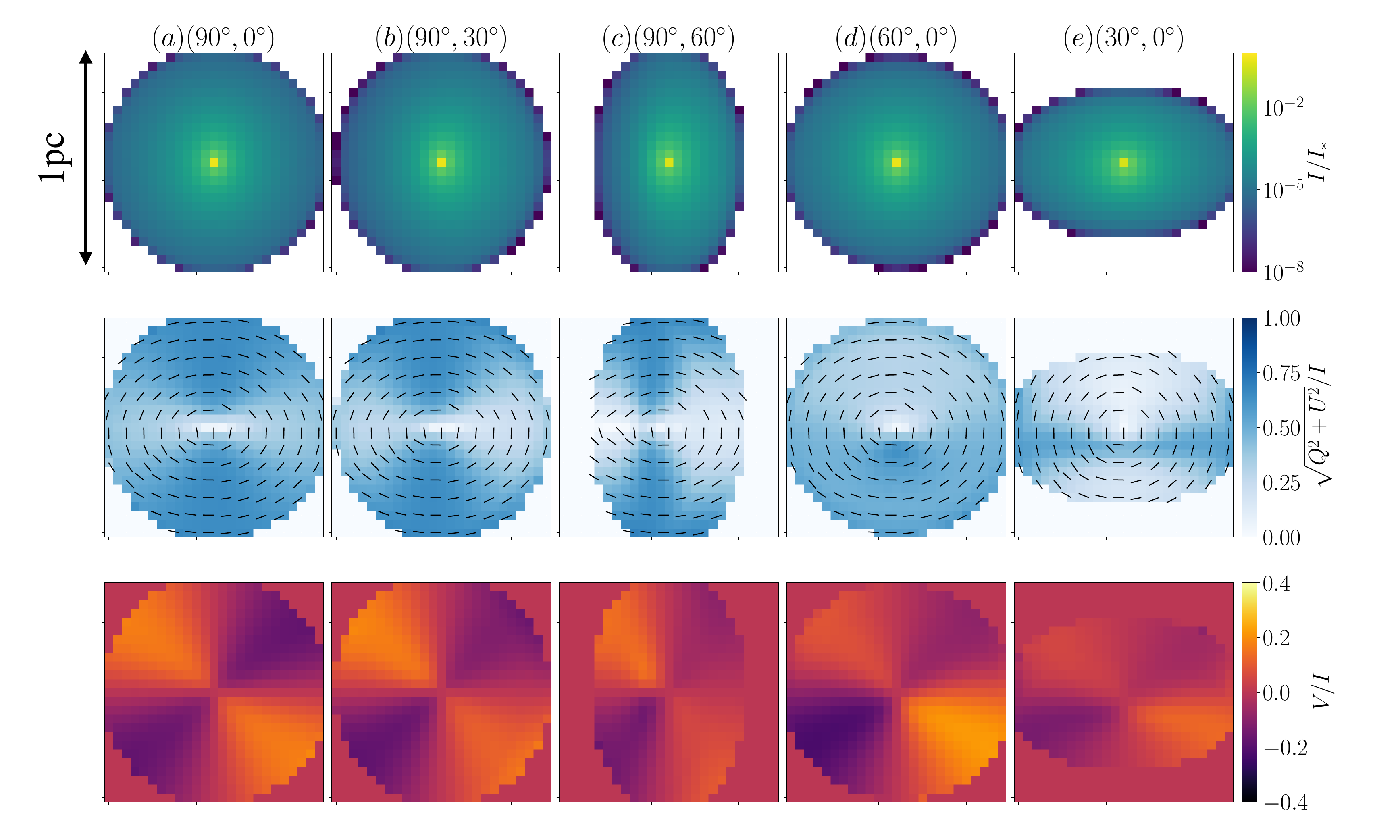}
	\end{center}
	\caption{
    The observational images of Ly$\alpha$ ($\lambda = 0.1216~{\mu {\rm m}}$).
	The angles $(\theta_{\rm obs}, \phi_{\rm obs})$ represent the positions of observers as shown in Fig.\ref{fig_slab}. 
	The top, middle, and bottom columns show the intensity distribution, linear polarization, and circular polarization.
	In the top panels, we plot the ratio of the observed light to stellar light.
	In the middle panels, the black bars show the direction of linear polarization.
	}
	\label{fig3}
\end{figure*}

\section{Results}\label{Sec:results}

\subsection{Observational maps}\label{Sect:Fiducial_MRN_mixture_model_NIR}

We first show the results on the polarization of Ly$\alpha$.
Fig. \ref{fig3} represents the observational images of the intensity and the degree of the linear/circular polarization at each viewing angle.
The intensity maps are normalized with the intrinsic intensity $I_*$ emitted from the radiation source. 
The intensity monotonically decreases as the radial distance increases.
The centre is the brightest because photons directly escape from the slab without scattering.

The middle panels in Fig. \ref{fig3} show the linear polarization degree and direction. 
The scattered light is linearly polarized along the perpendicular direction of the scattering plane, resulting in the concentric pattern shown in the figure.
The degree of linear polarization is smaller in the horizontal direction of the slab as shown in the face-on map (Fig.\ref{fig3}-a), and it lessens in the forward and backward scattered light as shown in Fig.\ref{fig3} (b)-(e).
The linear polarization degree of scattered light by large dust grains is lower than that of small grains
\citetalias{2020MNRAS.496.2762F}.
At the wavelength of Ly$\alpha$, the Rayleigh approximation is valid only at $a_{\rm d} < 10^{-2}~{\rm \mu m}$.
The large dust grains reduce linear polarization degrees.

\begin{figure}
	\begin{center}
	\includegraphics[width=\columnwidth]{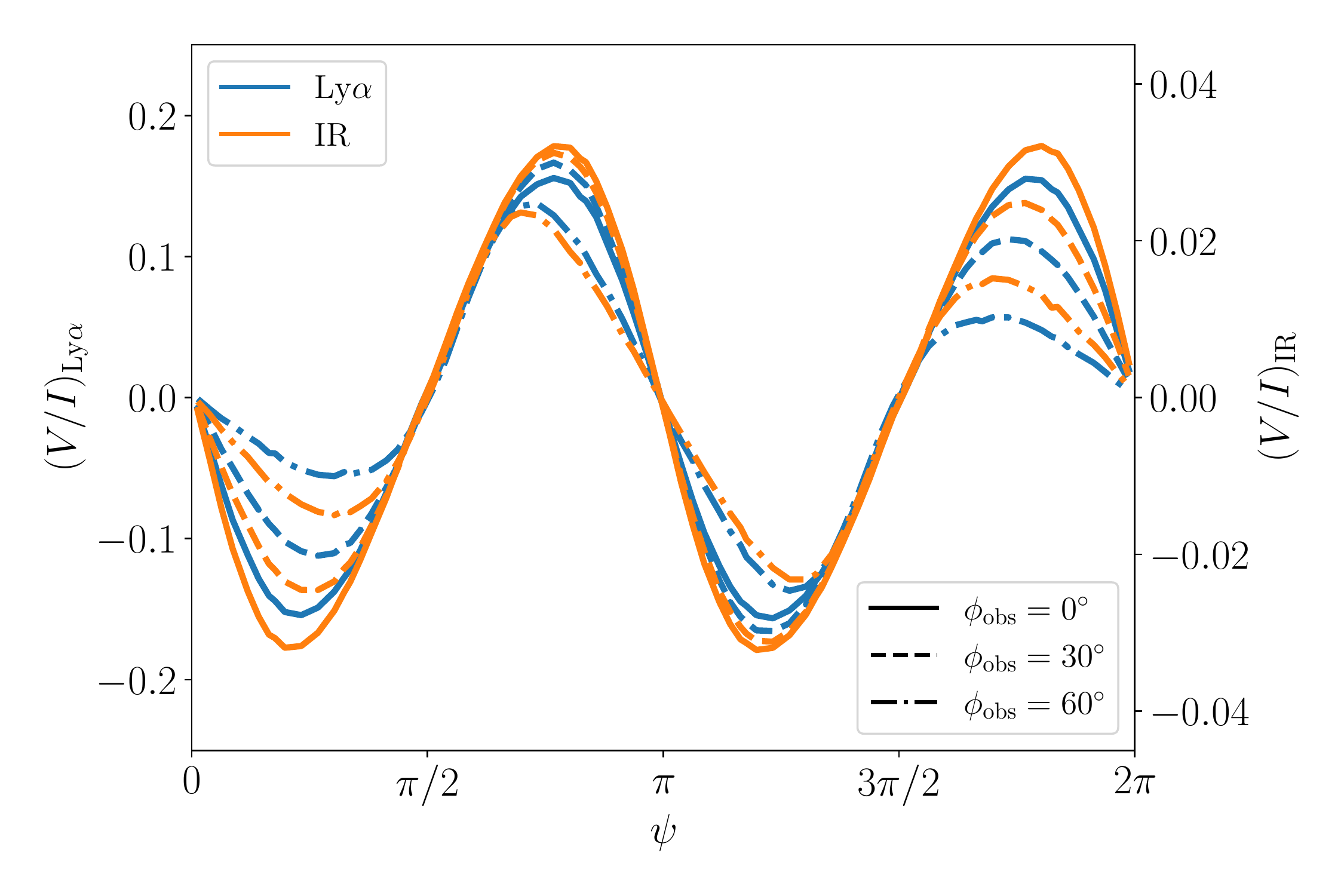}
	\end{center}
	\caption{
    The CP degree along the circle of radius $0.25 ~{\rm pc}$ at the centre of the observation maps.
    Each line shows the cases with the viewing angles $(\theta_{\rm obs}, \phi_{\rm obs}) = (90^{\circ}, 0^{\circ})$, $(90^{\circ}, 30^{\circ})$, and $(90^{\circ}, 60^{\circ})$, which correspond to the bottom panels of Fig. \ref{fig3}-a, b, and c.
	The horizontal axis is the rotational axis ($\psi$).
	Each line shows the cases of Ly$\alpha$ (blue) and the NIR (orange) wavelength.
	The left (right) vertical axis shows the CP degree of Ly$\alpha$ (the NIR wavelength).
	The direction of the horizontal axis corresponds to $\psi = 0^{\circ}$.
	} \label{fig_cpmax2}
\end{figure}

\begin{figure}
	\begin{center}
	\includegraphics[width=\columnwidth]{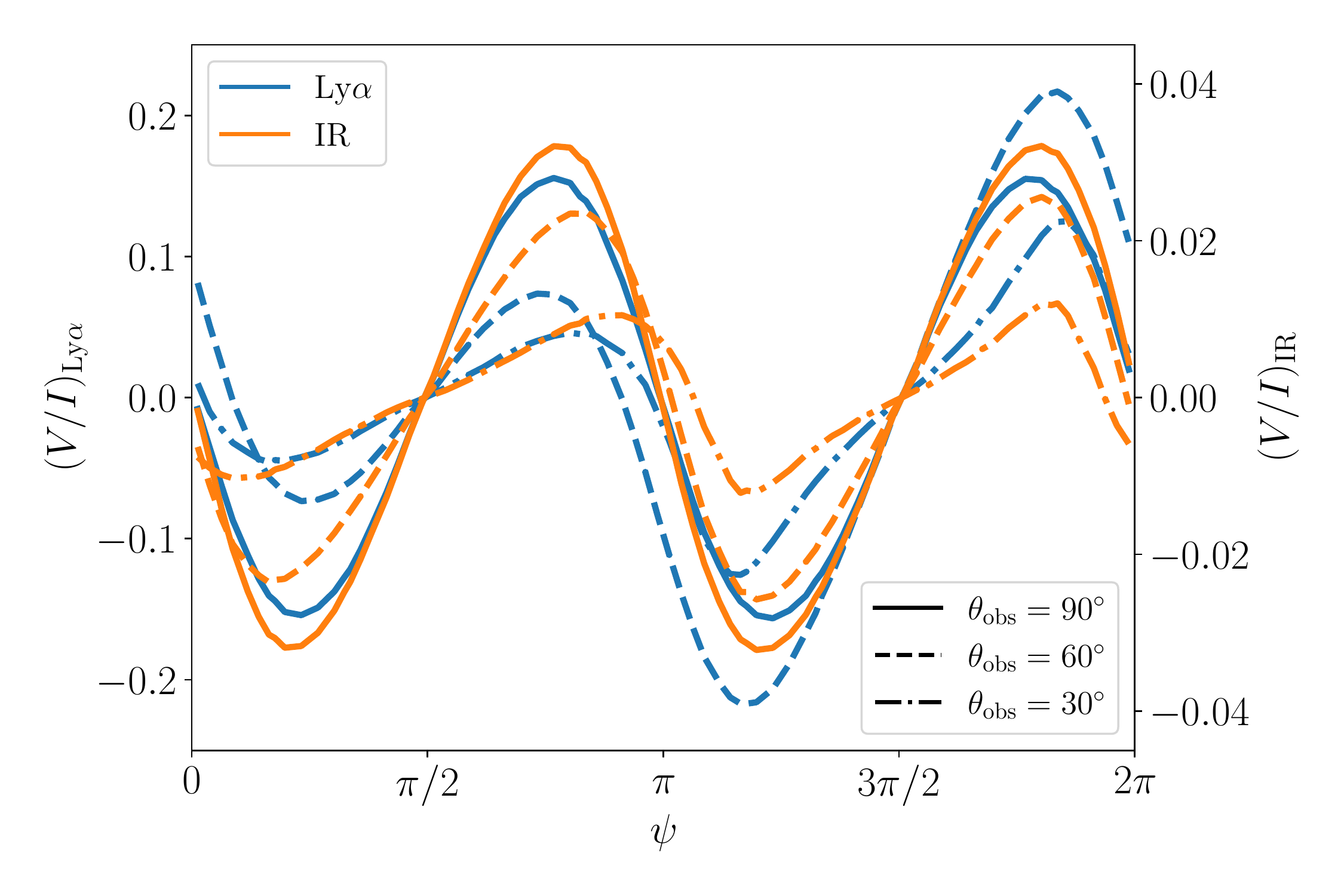}
	\end{center}
	\caption{Same as Fig. \ref{fig_cpmax2}, but for the observational angles $(\theta_{\rm obs}, \phi_{\rm obs}) = (90^{\circ}, 0^{\circ})$, $(60^\circ, 0^\circ)$, and $(30^\circ, 0^\circ)$, which correspond to the bottom panels of Fig. \ref{fig3}-a, d, and e. }
	\label{fig_cpmax3}
\end{figure}

The bottom panels of Fig. \ref{fig3} show the distributions of the CP degree.
We find that the symmetric quadrupole patterns appear 
like that at the NIR wavelength (see \citetalias{2020MNRAS.496.2762F}).
In Fig. \ref{fig_cpmax2} and \ref{fig_cpmax3}, we show the azimuthal distributions of the CP degree along the circle of the radius $0.25~{\rm pc}$ in Fig. \ref{fig3}.
In the face-on view (Fig. \ref{fig3}-a), the maximum value of the CP degree reaches $\sim 15$ percent.  
The relation between the CP degree and the inclination angle is not simple. 
In the case of $(\theta_{\rm obs}, \phi_{\rm obs}) = (60^\circ, 0^\circ)$, the fraction of the back scattering photons in the observable flux becomes large, resulting in the high CP degree $\sim 20$ percent. 
Fig. \ref{fig_cpmax2} and \ref{fig_cpmax3} also show the azimuthal distributions of the CP at the NIR wavelength as shown in Fig. \ref{fig2}.
The absolute values of CP degree are the largest ($|V/I| \sim 0.03$) in the diagonal direction of the face-on view (Fig. \ref{fig2}-a).
The CP degree decreases as the inclination angle increases because of the contributions of the forward and backward scattering photons.
The CP degree at the NIR wavelength is correlated with that of Ly$\alpha$ one and it drops by a factor $\sim 5$.

\subsection{Physical origin of high CP degree of Ly$\alpha$}\label{Sec:UV_CP_cal}

\begin{figure}
	\begin{center}
	\includegraphics[width=\columnwidth]{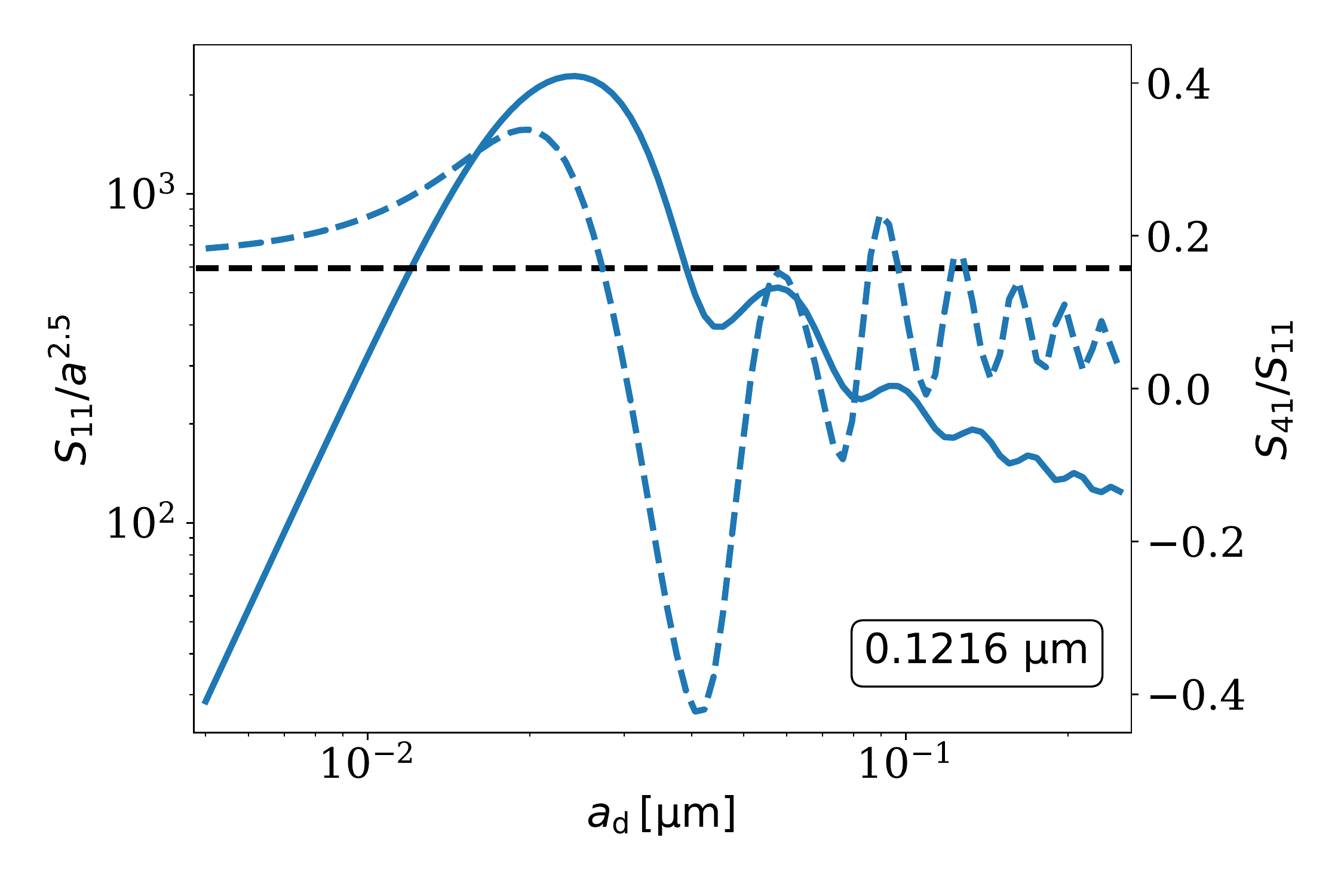}
	\end{center}
	\caption{
	The size dependence of the cross-section and CP degree in the cases with Ly$\alpha$.
	Each lines shows the cases of $(\Theta, \theta_{\rm s}, \phi_{\rm s}) = (45^{\circ}, 90^{\circ}, 270^{\circ})$.
	The solid line represents $S_{11}/a^{2.5}$.
	The dashed line shows $S_{41}/S_{11}$.
	The black dashed-line is averaged value of $S_{41}/S_{11}$ with the MRN mixture ($n_{\rm d} \propto a^{-3.5}$).
	}
	\label{fig:cp_sca_01216}
\end{figure}

\begin{figure}
	\begin{center}
	\includegraphics[width=\columnwidth]{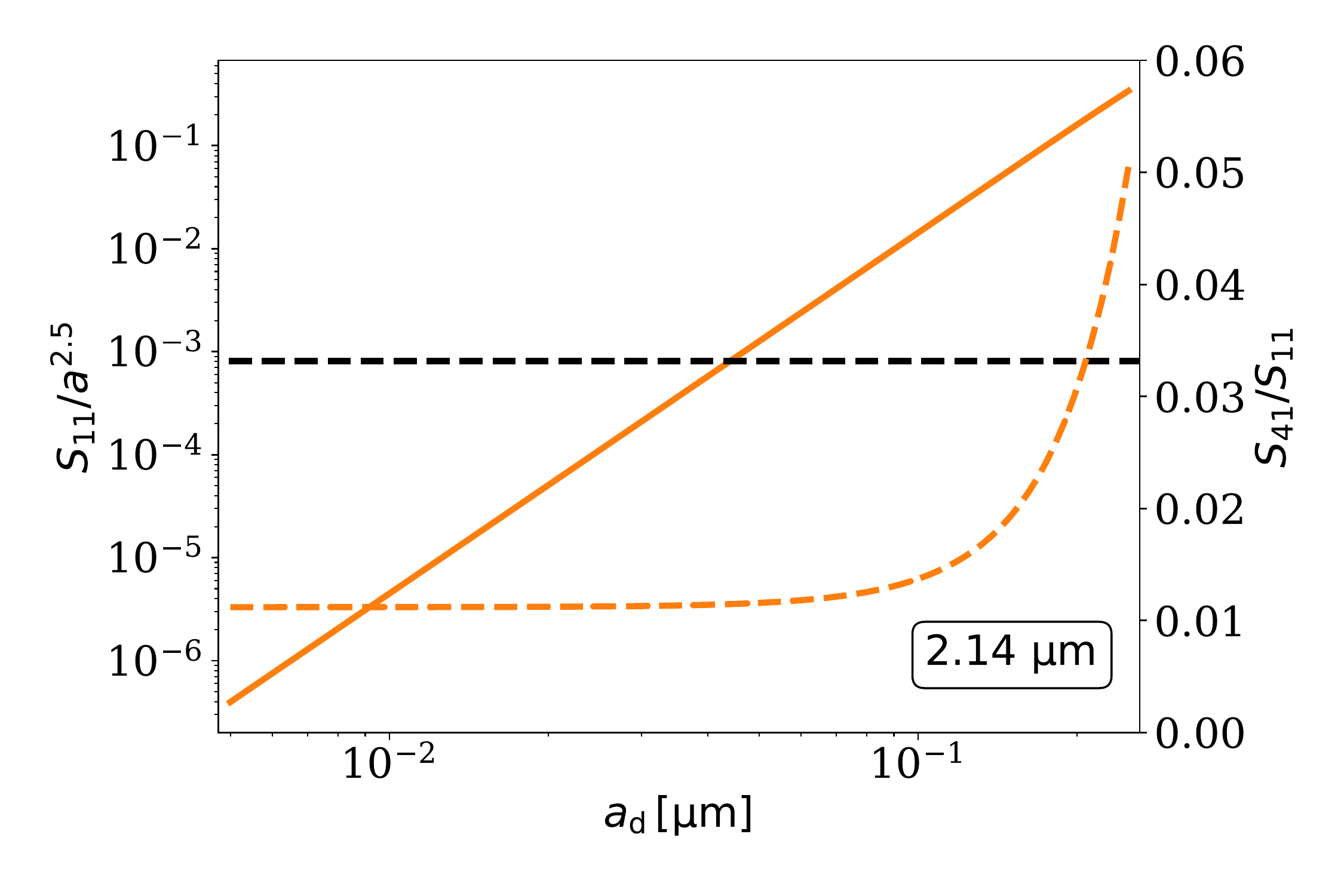}
	\end{center}
 	\caption{
	Same as Fig. \ref{fig:cp_sca_01216}, but for the case with $\lambda = 2.14~{\mu \rm m}$.}
	\label{fig:cp_sca_214}
\end{figure}

\begin{figure}
	\begin{center}
	\includegraphics[width=\columnwidth]{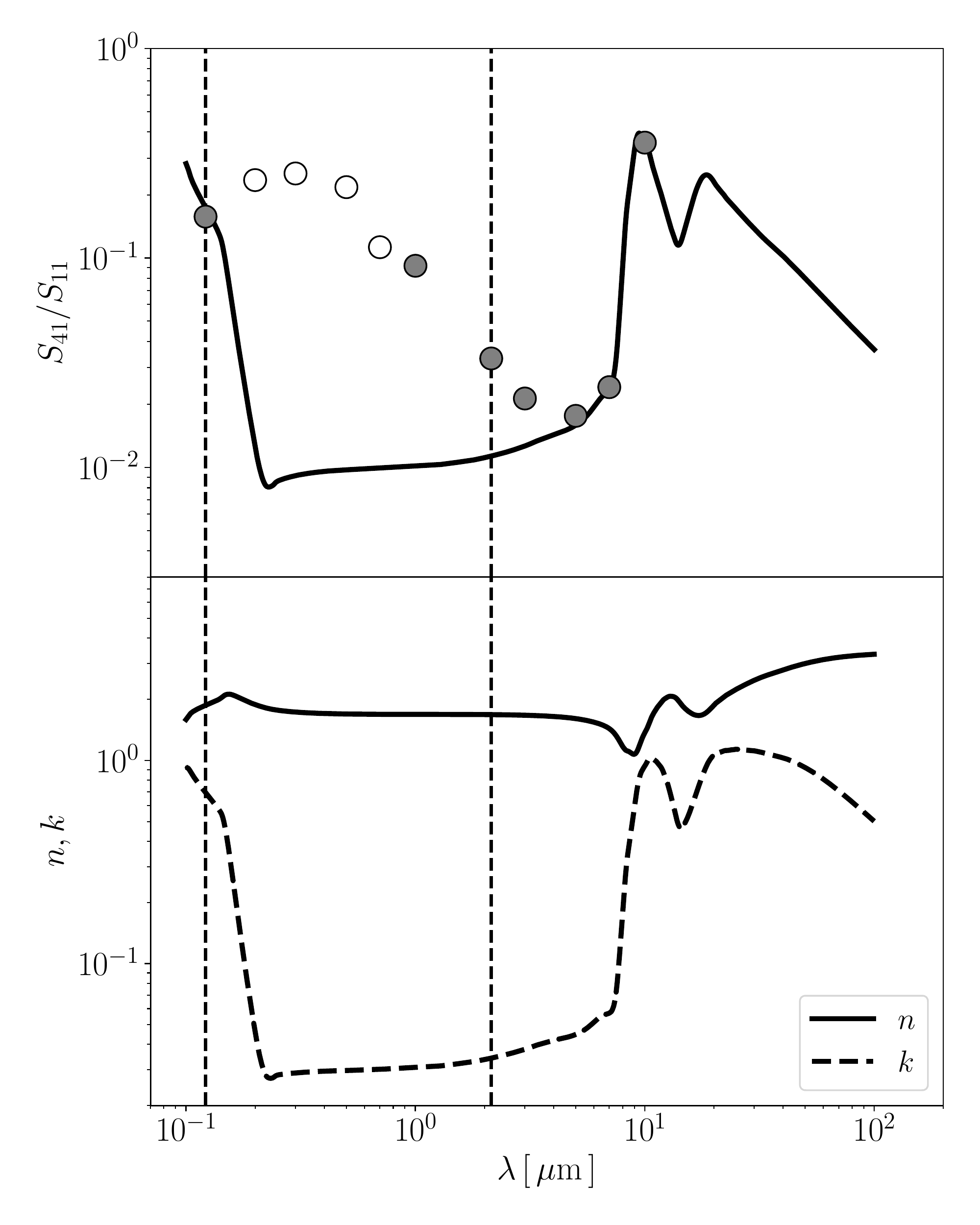}
	\end{center}
	\caption{
	Upper panel: the dependence of $S_{41}/S_{11}$ on the wavelength in the case of the scattering angles $(\Theta, \theta_{\rm s}, \phi_{\rm s}) = (45^\circ, 90^\circ, 270^\circ)$. The circles show the mean absolute values 
	calculated by {\sc DDSCAT} code. The filled and non-filled circles represent the positive and negative values. The solid line shows the values calculated with Equations \ref{eq_S11} and \ref{eq_S41} in the Rayleigh approximation. The vertical dashed lines show the wavelength considered in this study ($\lambda = 0.1216~{\rm \mu m}$ and $2.14~{\rm \mu m}$). Bottom panel: the dependence of the refractive index of astronomical silicate \citep{2003ApJ...598.1017D}.
	}
	\label{fig:s41s11_astrosil}
\end{figure}

Here we investigate the origin of high CP degree of Ly$\alpha$.
The scattering cross-section for non-polarized photons is determined solely by the $S_{11}$ component of the M\"uller matrix (see also Eq. \ref{cross_section_sca}).
The abundance of dust grains in the logarithmic bin is given as $\propto a_{\rm d}^{-2.5}$.
Thus, the values of $S_{11}/a_{\rm d}^{2.5}$ represent the weight of each logarithmic bin on the cross-section.
Fig. \ref{fig:cp_sca_01216} shows the size dependence of $S_{11}/a_{\rm d}^{2.5}$ in the case with $(\Theta, \theta_{\rm s}, \phi_{\rm s}) = (45^{\circ}, 90^{\circ}, 270^{\circ})$ at the wavelength of Ly$\alpha$.
At this angle, the CP degree becomes maximum if the Rayleigh approximation is valid (see also Eq. \ref{eq_S41}).
In the Rayleigh approximation, the cross-section for scattering is proportional to the dust grain size as $\propto a_{\rm d}^6$, resulting in that $S_{11}/a_{\rm d}^{2.5} \propto a_{\rm d}^{3.5}$.
At $a_{\rm d} < 0.025~{\rm \mu m}$, the value of $S_{11}/a_{\rm d}^{2.5}$ monotonically increases as the Rayleigh approximation.
At $a_{\rm d} > 0.025~{\rm \mu m}$, $S_{11}/a_{\rm d}^{2.5}$ oscillates and damps.
Thus, the dust grains of $a_{\rm d} \sim 0.025~{\rm \mu m}$ mainly contribute to the scattering process.
The Rayleigh approximation for $S_{41}/S_{11}$ is valid only if $a_{\rm d} < 10^{-2}~{\rm \mu m}$.
It slightly increases to $\sim 0.3$ at $\sim 0.02~{\rm \mu m}$, but it rapidly decreases to $-0.4$ at  $a_{\rm d} \sim 0.04~{\rm \mu m}$.
At the larger dust grains, $S_{41}/S_{11}$ fluctuates and converges to zero at $a_{\rm d} \gtrsim 0.2~\rm \mu m$. 
The averaged value of $S_{41}/S_{11}$ is $\sim 0.2$, which is close to that of the Rayleigh approximation.

Fig. \ref{fig:cp_sca_214} shows the dust size dependence of $S_{11}/a_{\rm d}^{2.5}$ and $S_{41}/S_{11}$ at the NIR wavelength to compare that of the wavelength of  Ly$\alpha$.
The value of $S_{11}/a_{\rm d}^{2.5}$ monotonically increases as $\propto a_{\rm d}^{3.5}$ as well as the Rayleigh approximation.
Therefore, the larger dust grains have more contributions to the scattering processes.
The values of $S_{41}/S_{11}$ is constant at $a_{\rm d}<10^{-1}~{\rm \mu m}$ where the Rayleigh approximation is valid (also see Appendix \ref{apdB}).
For the larger dust grains, $S_{41}/S_{11}$ increases up to 0.05 at $a_{\rm d}=0.25~{\rm \mu m}$.
These large dust grains raise the averaged value of $S_{41}/S_{11}$ in the case of the MRN size distribution model at the NIR wavelength.

The trend of $S_{41}/S_{11}$ at each wavelength can be understood from the view of the Rayleigh approximation.
Here, we compare our models based on DDSCAT with the Rayleigh approximation.
The solid line in the top panel of Fig. \ref{fig:s41s11_astrosil} shows the $S_{41}/S_{11}$ calculated based on the Rayleigh approximation and the refractive index (see also Appendix \ref{apdB}).
The value of $S_{41}/S_{11}$ increases with the ratio of the imaginary part to the real part of the refractive index ($m=n+ik$).
In this figure, we assume that dust grains are oblate whose axis ratio is $q = 0.5$, and the scattering angles are $(\Theta, \theta_{\rm s}, \phi_{\rm s}) = (45^\circ, 90^\circ, 270^\circ)$.
The circle symbols in Fig. \ref{fig:s41s11_astrosil} show the mean values of $S_{41}/S_{11}$ calculated with {\sc DDSCAT} code.
At $\lambda > 3~{\rm \mu m}$, the value of $S_{41}/S_{11}$ nicely match the values derived from the Rayleigh approximation.
At the shorter wavelength, the value of $S_{41}/S_{11}$ deviates from the Rayleigh approximation and increases up to 10 percent due to the contribution of larger dust grains as shown in Fig. \ref{fig:cp_sca_214}.
Also, the sign of $S_{41}/S_{11}$ is reversed at $0.2~{\rm \mu m} < \lambda < 1~{\rm \mu m}$ because the large dust grains have the negative values in these wavelengths.
At $\lambda < 0.2~{\rm \mu m}$, the imaginary part of the refractive index increases, and the fluctuation of $S_{41}/S_{11}$ occurs as shown in Fig. \ref{fig:cp_sca_01216}.
Due to these oscillations of $S_{41}/S_{11}$,
the contributions of the large dust grains are cancelled out.
Thus, the mean value of $S_{41}/S_{11}$ reaches the Rayleigh approximation at $\lambda = 0.1216~{\rm \mu m}$.

\section{Summary and Discussion}\label{sec:summary_discussion}

We have studied the generation of CP via multiple scattering processes on dust grains by performing three-dimensional radiative transfer simulations.
We have investigated the CP degree of Ly$\alpha$ 
($\lambda=0.1216{\rm \mu m}$) and NIR wavelength ($\lambda=2.14{\rm \mu m}$), assuming a dusty slab composed of aligned dust grains.
The quadrupole patterns of CP appear in both cases, regardless of the viewing angles.
We find that the CP of Ly$\alpha$ can become much higher than that at the NIR wavelength.
The CP degree is $|V|/I \sim 15$ percent for Ly$\alpha$ and $|V|/I \sim 3$ percent for the NIR wavelength.

\citet{2005OLEB...35...29L} suggested that the high CP degree at the UV wavelengths ($\lambda > 0.2 ~\rm \mu m$) was generated by the dichroic extinction process.
In this work, we have focused on the shorter wavelength $\lambda=0.1216~\rm \mu m$, corresponding to Ly$\alpha$ line of hydrogen.
The Ly$\alpha$ photons are emitted from the ionized gas associated with star-forming regions.
The circularly polarized Ly$\alpha$ photons can cause the enantiomeric excess of amino acids \citep{shoji2023enantiomeric}
and chiral molecules \citep{hori2022theoretical}.
Therefore, interstellar Ly$\alpha$ radiation with high CP as shown in our models might induce the enantiomeric excess of amino acids discovered in the meteorites \citep{1997Natur.389..265E, 2018P&SS..164..127P}.

In this study, we have supposed a dusty slab
composed of oblate dust grains whose minor axes are aligned with the edge-on angle of the slab.
The optical depth of the slab is assumed to be unity
in the perpendicular direction. 
In such a configration, single-scattered photons mostly contribute to the CP generation. 
Although we have adopted a simplified model of a dusty slab,
the observed quadruple structure of the CP at near-infrared wavelengths is well reproduced. 
In optically thick cases, the multiply scattered photons also contribute to the CP generation via the dichroic extinction process \citep{2005OLEB...35...29L}. 
Actually, the distributions and aligned directions of dust grains depend on the gas streaming, magnetic, and radiation fields \citep[][and reference therein]{2015ARA&A..53..501A}.
However, the spatial distributions of gas and magnetic fields are still under debate because of complicated processes with stellar feedback \citep[e.g.,][]{2018ApJ...859...68K, 2021ApJ...911..128K, 2020MNRAS.497.3830F}. 
In a future work, we will perform simulations using the results of the magneto-radiation hydrodynamics simulations to investigate the effects of multiple scattering and the alignment of dust grains in star-forming clouds.

In this work, we have considered the astronomical silicate dust model alone. On the other hand, the two-component model with silicate and graphite has been taken into account frequently in previous works \citep[e.g.][]{1977ApJ...217..425M, 1984ApJ...285...89D}.
Since the Stokes parameters depend on the dust component, the CP degrees in our simulations can change with different dust models. 
Although it is difficult to constrain the dust properties only from the observational data,
\citet{2021ApJ...909...94D} recently found that the single component model of large dust grains $(\gtrsim 0.01~\mu {\rm m})$ 
can reproduce the observed extinction well \citep[see also][]{2022arXiv220812365H}.
In their model, the dielectric function of grains at $h \nu \gtrsim 8 ~{\rm eV}$ is almost the same as the model used in this study.

We have studied the interaction between photons and dust grains. However, Ly$\alpha$ photons also interact with neutral hydrogen atoms \citep{2012MNRAS.424..884Y, 2018MNRAS.476.2664A} that can induce the linear polarization \citep{2022ApJS..259....3S}. Then, the linear polarized light can be converted into the circularly polarized one via scattering with a dust grain or a hydrogen atom. 
On the other hand, the coherency of the last scattering angle can get lower by multiple scattering with hydrogen atoms and dust grains, which likely decreases the CP degree.

The observation of the polarized light is a powerful tool to investigate physical properties in the interstellar medium, such as the magnetic fields \citep[e.g.,][]{1949Sci...109..166H, 1949ApJ...109..471H}.
Recently, the observation of Atacama Large Millimeter/submillimeter Array (ALMA) discovered the linearly polarized light in the protoplanetary disks, and it is mainly caused by the scattering on the dust grains \citep[e.g.,][]{2015ApJ...809...78K, 2017ApJ...839...56T}.  
The observations of the circularly polarized light will add more physical information about the interstellar medium and the protoplanetary disks \citep{2022ApJ...926...90D}.
CP light has been mainly observed in the near-infrared \citep[$1~{\rm \mu m} \lesssim \lambda \lesssim 3~{\rm \mu m}$, e.g.,][]{2014ApJ...795L..16K}, but we expect that the degree of CP becomes higher at the mid-infrared (MIR) wavelength ($10~{\rm \mu m} \lesssim \lambda \lesssim 30~{\rm \mu m}$) as shown in Fig. \ref{fig:s41s11_astrosil}.
In particular, the emission spectrum of an embedded massive protostar has a peak around the mid-infrared \citep{2016ApJ...818...52T, 2018MNRAS.473.4754F}.
Therefore, highly circularly polarized light in the mid-infrared wavelength is expected to exist around forming massive stars.
Note that, on the other hand, thermal emission from hot dust can overwhelm the scattered light at MIR wavelengths in regions where massive stars distribute densely. If the thermal emission directly escapes the system, the CP degrees become lower.

Circularly polarized UV can induce
the enantiomeric excess of the biological molecules
in meteorites, which fell on to the earth and might
trigger the homochirality. 
In this paper, we have demonstrated that the Ly$\alpha$ CP is tightly correlated with that at the infrared wavelength regardless of the viewing angles and is higher by a factor of $\sim 5$ than that at the infrared wavelength.
We suggest that interstellar Ly$\alpha$ radiation with high CP in early evolutionary phase of
the Milky Way might initialize enantiomeric excess of the biological molecules.

\section*{Acknowledgements}
This work is supported in part by MEXT/JSPS KAKENHI Grant Number JP23K13139(HF), JP17H04827, JP20H04724, JP21H04489 (HY), JP19H00697(MU), NAOJ ALMA Scientific Research Grant Numbers 2019-11A, JST FOREST Program, Grant Number JP-MJFR202Z (HY), and Astro Biology Center Project research AB041008 (HY). 

\section*{DATA AVAILABILITY}
The data underlying this article will be shared on reasonable request to the corresponding author.



\bibliographystyle{mnras}

\begin{thebibliography}{}
\makeatletter
\relax
\def\mn@urlcharsother{\let\do\@makeother \do\$\do\&\do\#\do\^\do\_\do\%\do\~}
\def\mn@doi{\begingroup\mn@urlcharsother \@ifnextchar [ {\mn@doi@}
  {\mn@doi@[]}}
\def\mn@doi@[#1]#2{\def\@tempa{#1}\ifx\@tempa\@empty \href
  {http://dx.doi.org/#2} {doi:#2}\else \href {http://dx.doi.org/#2} {#1}\fi
  \endgroup}
\def\mn@eprint#1#2{\mn@eprint@#1:#2::\@nil}
\def\mn@eprint@arXiv#1{\href {http://arxiv.org/abs/#1} {{\tt arXiv:#1}}}
\def\mn@eprint@dblp#1{\href {http://dblp.uni-trier.de/rec/bibtex/#1.xml}
  {dblp:#1}}
\def\mn@eprint@#1:#2:#3:#4\@nil{\def\@tempa {#1}\def\@tempb {#2}\def\@tempc
  {#3}\ifx \@tempc \@empty \let \@tempc \@tempb \let \@tempb \@tempa \fi \ifx
  \@tempb \@empty \def\@tempb {arXiv}\fi \@ifundefined
  {mn@eprint@\@tempb}{\@tempb:\@tempc}{\expandafter \expandafter \csname
  mn@eprint@\@tempb\endcsname \expandafter{\@tempc}}}

\bibitem[\protect\citeauthoryear{{Abe}, {Suzuki}, {Hasegawa}, {Semelin},
  {Yajima}  \& {Umemura}}{{Abe} et~al.}{2018}]{2018MNRAS.476.2664A}
{Abe} M.,  {Suzuki} H.,  {Hasegawa} K.,  {Semelin} B.,  {Yajima} H.,
  {Umemura} M.,  2018, \mn@doi [\mnras] {10.1093/mnras/sty233}, \href
  {https://ui.adsabs.harvard.edu/abs/2018MNRAS.476.2664A} {476, 2664}

\bibitem[\protect\citeauthoryear{{Andersson}, {Lazarian}  \&
  {Vaillancourt}}{{Andersson} et~al.}{2015}]{2015ARA&A..53..501A}
{Andersson} B.~G.,  {Lazarian} A.,   {Vaillancourt} J.~E.,  2015, \mn@doi
  [\araa] {10.1146/annurev-astro-082214-122414}, \href
  {https://ui.adsabs.harvard.edu/abs/2015ARA&A..53..501A} {53, 501}

\bibitem[\protect\citeauthoryear{{Arzoumanian} et~al.,}{{Arzoumanian}
  et~al.}{2011}]{2011A&A...529L...6A}
{Arzoumanian} D.,  et~al., 2011, \mn@doi [\aap] {10.1051/0004-6361/201116596},
  \href {https://ui.adsabs.harvard.edu/abs/2011A&A...529L...6A} {529, L6}

\bibitem[\protect\citeauthoryear{{Bailey}, {Chrysostomou}, {Hough}, {Gledhill},
  {McCall}, {Clark}, {Menard}  \& {Tamura}}{{Bailey}
  et~al.}{1998}]{1998Sci...281..672B}
{Bailey} J.,  {Chrysostomou} A.,  {Hough} J.~H.,  {Gledhill} T.~M.,  {McCall}
  A.,  {Clark} S.,  {Menard} F.,   {Tamura} M.,  1998, \mn@doi [Science]
  {10.1126/science.281.5377.672}, \href
  {https://ui.adsabs.harvard.edu/abs/1998Sci...281..672B} {281, 672}

\bibitem[\protect\citeauthoryear{{Bohren} \& {Huffman}}{{Bohren} \&
  {Huffman}}{1983}]{1983asls.book.....B}
{Bohren} C.~F.,  {Huffman} D.~R.,  1983, {Absorption and scattering of light by
  small particles}

\bibitem[\protect\citeauthoryear{{Bonner}}{{Bonner}}{1991}]{1991OLEB...21...59B}
{Bonner} W.~A.,  1991, \mn@doi [Origins of Life and Evolution of the Biosphere]
  {10.1007/BF01809580}, \href
  {https://ui.adsabs.harvard.edu/abs/1991OLEB...21...59B} {21, 59}

\bibitem[\protect\citeauthoryear{{Chrysostomou}, {Gledhill}, {M{\'e}nard},
  {Hough}, {Tamura}  \& {Bailey}}{{Chrysostomou}
  et~al.}{2000}]{2000MNRAS.312..103C}
{Chrysostomou} A.,  {Gledhill} T.~M.,  {M{\'e}nard} F.,  {Hough} J.~H.,
  {Tamura} M.,   {Bailey} J.,  2000, \mn@doi [\mnras]
  {10.1046/j.1365-8711.2000.03126.x}, \href
  {https://ui.adsabs.harvard.edu/abs/2000MNRAS.312..103C} {312, 103}

\bibitem[\protect\citeauthoryear{Condon}{Condon}{1937}]{RevModPhys.9.432}
Condon E.~U.,  1937, \mn@doi [Rev. Mod. Phys.] {10.1103/RevModPhys.9.432}, 9,
  432

\bibitem[\protect\citeauthoryear{{Draine}}{{Draine}}{2003}]{2003ApJ...598.1017D}
{Draine} B.~T.,  2003, \mn@doi [\apj] {10.1086/379118}, \href
  {https://ui.adsabs.harvard.edu/abs/2003ApJ...598.1017D} {598, 1017}

\bibitem[\protect\citeauthoryear{{Draine}}{{Draine}}{2022}]{2022ApJ...926...90D}
{Draine} B.~T.,  2022, \mn@doi [\apj] {10.3847/1538-4357/ac3977}, \href
  {https://ui.adsabs.harvard.edu/abs/2022ApJ...926...90D} {926, 90}

\bibitem[\protect\citeauthoryear{{Draine} \& {Flatau}}{{Draine} \&
  {Flatau}}{2013}]{2013arXiv1305.6497D}
{Draine} B.~T.,  {Flatau} P.~J.,  2013, arXiv e-prints, \href
  {https://ui.adsabs.harvard.edu/abs/2013arXiv1305.6497D} {p. arXiv:1305.6497}

\bibitem[\protect\citeauthoryear{{Draine} \& {Hensley}}{{Draine} \&
  {Hensley}}{2021}]{2021ApJ...909...94D}
{Draine} B.~T.,  {Hensley} B.~S.,  2021, \mn@doi [\apj]
  {10.3847/1538-4357/abd6c6}, \href
  {https://ui.adsabs.harvard.edu/abs/2021ApJ...909...94D} {909, 94}

\bibitem[\protect\citeauthoryear{{Draine} \& {Lee}}{{Draine} \&
  {Lee}}{1984}]{1984ApJ...285...89D}
{Draine} B.~T.,  {Lee} H.~M.,  1984, \mn@doi [\apj] {10.1086/162480}, \href
  {https://ui.adsabs.harvard.edu/abs/1984ApJ...285...89D} {285, 89}

\bibitem[\protect\citeauthoryear{{Engel} \& {Macko}}{{Engel} \&
  {Macko}}{1997}]{1997Natur.389..265E}
{Engel} M.~H.,  {Macko} S.~A.,  1997, \mn@doi [\nat] {10.1038/38460}, \href
  {https://ui.adsabs.harvard.edu/abs/1997Natur.389..265E} {389, 265}

\bibitem[\protect\citeauthoryear{{Fukue} et~al.,}{{Fukue}
  et~al.}{2010}]{2010OLEB...40..335F}
{Fukue} T.,  et~al., 2010, \mn@doi [Origins of Life and Evolution of the
  Biosphere] {10.1007/s11084-010-9206-1}, \href
  {https://ui.adsabs.harvard.edu/abs/2010OLEB...40..335F} {40, 335}

\bibitem[\protect\citeauthoryear{{Fukushima} \& {Yajima}}{{Fukushima} \&
  {Yajima}}{2021}]{2021MNRAS.506.5512F}
{Fukushima} H.,  {Yajima} H.,  2021, \mn@doi [\mnras] {10.1093/mnras/stab2099},
  \href {https://ui.adsabs.harvard.edu/abs/2021MNRAS.506.5512F} {506, 5512}

\bibitem[\protect\citeauthoryear{{Fukushima} \& {Yajima}}{{Fukushima} \&
  {Yajima}}{2022}]{2022MNRAS.511.3346F}
{Fukushima} H.,  {Yajima} H.,  2022, \mn@doi [\mnras] {10.1093/mnras/stac244},
  \href {https://ui.adsabs.harvard.edu/abs/2022MNRAS.511.3346F} {511, 3346}

\bibitem[\protect\citeauthoryear{{Fukushima}, {Omukai}  \&
  {Hosokawa}}{{Fukushima} et~al.}{2018}]{2018MNRAS.473.4754F}
{Fukushima} H.,  {Omukai} K.,   {Hosokawa} T.,  2018, \mn@doi [\mnras]
  {10.1093/mnras/stx2620}, \href
  {https://ui.adsabs.harvard.edu/abs/2018MNRAS.473.4754F} {473, 4754}

\bibitem[\protect\citeauthoryear{{Fukushima}, {Yajima}  \&
  {Umemura}}{{Fukushima} et~al.}{2020a}]{2020MNRAS.496.2762F}
{Fukushima} H.,  {Yajima} H.,   {Umemura} M.,  2020a, \mn@doi [\mnras]
  {10.1093/mnras/staa1718}, \href
  {https://ui.adsabs.harvard.edu/abs/2020MNRAS.496.2762F} {496, 2762}

\bibitem[\protect\citeauthoryear{{Fukushima}, {Yajima}, {Sugimura}, {Hosokawa},
  {Omukai}  \& {Matsumoto}}{{Fukushima} et~al.}{2020b}]{2020MNRAS.497.3830F}
{Fukushima} H.,  {Yajima} H.,  {Sugimura} K.,  {Hosokawa} T.,  {Omukai} K.,
  {Matsumoto} T.,  2020b, \mn@doi [\mnras] {10.1093/mnras/staa2062}, \href
  {https://ui.adsabs.harvard.edu/abs/2020MNRAS.497.3830F} {497, 3830}

\bibitem[\protect\citeauthoryear{{Gledhill} \& {McCall}}{{Gledhill} \&
  {McCall}}{2000}]{2000MNRAS.314..123G}
{Gledhill} T.~M.,  {McCall} A.,  2000, \mn@doi [\mnras]
  {10.1046/j.1365-8711.2000.03323.x}, \href
  {https://ui.adsabs.harvard.edu/abs/2000MNRAS.314..123G} {314, 123}

\bibitem[\protect\citeauthoryear{{G{\'o}rski}, {Hivon}, {Banday}, {Wand elt},
  {Hansen}, {Reinecke}  \& {Bartelmann}}{{G{\'o}rski}
  et~al.}{2005}]{2005ApJ...622..759G}
{G{\'o}rski} K.~M.,  {Hivon} E.,  {Banday} A.~J.,  {Wand elt} B.~D.,  {Hansen}
  F.~K.,  {Reinecke} M.,   {Bartelmann} M.,  2005, \mn@doi [\apj]
  {10.1086/427976}, \href
  {https://ui.adsabs.harvard.edu/abs/2005ApJ...622..759G} {622, 759}

\bibitem[\protect\citeauthoryear{{Hall}}{{Hall}}{1949}]{1949Sci...109..166H}
{Hall} J.~S.,  1949, \mn@doi [Science] {10.1126/science.109.2825.166}, \href
  {https://ui.adsabs.harvard.edu/abs/1949Sci...109..166H} {109, 166}

\bibitem[\protect\citeauthoryear{{Hensley} \& {Draine}}{{Hensley} \&
  {Draine}}{2022}]{2022arXiv220812365H}
{Hensley} B.~S.,  {Draine} B.~T.,  2022, \mn@doi [arXiv e-prints]
  {10.48550/arXiv.2208.12365}, \href
  {https://ui.adsabs.harvard.edu/abs/2022arXiv220812365H} {p. arXiv:2208.12365}

\bibitem[\protect\citeauthoryear{{Hiltner}}{{Hiltner}}{1949}]{1949ApJ...109..471H}
{Hiltner} W.~A.,  1949, \mn@doi [\apj] {10.1086/145151}, \href
  {https://ui.adsabs.harvard.edu/abs/1949ApJ...109..471H} {109, 471}

\bibitem[\protect\citeauthoryear{Hori, Nakamura, Sakawa, Watanabe, Kayanuma,
  Shoji, Umemura  \& Shigeta}{Hori et~al.}{2022}]{hori2022theoretical}
Hori Y.,  Nakamura H.,  Sakawa T.,  Watanabe N.,  Kayanuma M.,  Shoji M.,
  Umemura M.,   Shigeta Y.,  2022, Astrobiology

\bibitem[\protect\citeauthoryear{{Kataoka} et~al.,}{{Kataoka}
  et~al.}{2015}]{2015ApJ...809...78K}
{Kataoka} A.,  et~al., 2015, \mn@doi [\apj] {10.1088/0004-637X/809/1/78}, \href
  {https://ui.adsabs.harvard.edu/abs/2015ApJ...809...78K} {809, 78}

\bibitem[\protect\citeauthoryear{{Kim}, {Kim}  \& {Ostriker}}{{Kim}
  et~al.}{2018}]{2018ApJ...859...68K}
{Kim} J.-G.,  {Kim} W.-T.,   {Ostriker} E.~C.,  2018, \mn@doi [\apj]
  {10.3847/1538-4357/aabe27}, \href
  {https://ui.adsabs.harvard.edu/abs/2018ApJ...859...68K} {859, 68}

\bibitem[\protect\citeauthoryear{{Kim}, {Ostriker}  \& {Filippova}}{{Kim}
  et~al.}{2021}]{2021ApJ...911..128K}
{Kim} J.-G.,  {Ostriker} E.~C.,   {Filippova} N.,  2021, \mn@doi [\apj]
  {10.3847/1538-4357/abe934}, \href
  {https://ui.adsabs.harvard.edu/abs/2021ApJ...911..128K} {911, 128}

\bibitem[\protect\citeauthoryear{Kuhn}{Kuhn}{1930}]{kuhn1930physical}
Kuhn W.,  1930, Transactions of the Faraday Society, 26, 293

\bibitem[\protect\citeauthoryear{{Kwon} et~al.,}{{Kwon}
  et~al.}{2013}]{2013ApJ...765L...6K}
{Kwon} J.,  et~al., 2013, \mn@doi [\apjl] {10.1088/2041-8205/765/1/L6}, \href
  {https://ui.adsabs.harvard.edu/abs/2013ApJ...765L...6K} {765, L6}

\bibitem[\protect\citeauthoryear{{Kwon} et~al.,}{{Kwon}
  et~al.}{2014}]{2014ApJ...795L..16K}
{Kwon} J.,  et~al., 2014, \mn@doi [\apjl] {10.1088/2041-8205/795/1/L16}, \href
  {https://ui.adsabs.harvard.edu/abs/2014ApJ...795L..16K} {795, L16}

\bibitem[\protect\citeauthoryear{{Kwon}, {Tamura}, {Hough}, {Nagata}  \&
  {Kusakabe}}{{Kwon} et~al.}{2016}]{2016AJ....152...67K}
{Kwon} J.,  {Tamura} M.,  {Hough} J.~H.,  {Nagata} T.,   {Kusakabe} N.,  2016,
  \mn@doi [\aj] {10.3847/0004-6256/152/3/67}, \href
  {https://ui.adsabs.harvard.edu/abs/2016AJ....152...67K} {152, 67}

\bibitem[\protect\citeauthoryear{{Kwon} et~al.,}{{Kwon}
  et~al.}{2018}]{2018AJ....156....1K}
{Kwon} J.,  et~al., 2018, \mn@doi [\aj] {10.3847/1538-3881/aac389}, \href
  {https://ui.adsabs.harvard.edu/abs/2018AJ....156....1K} {156, 1}

\bibitem[\protect\citeauthoryear{{Lada} \& {Lada}}{{Lada} \&
  {Lada}}{2003}]{2003ARA&A..41...57L}
{Lada} C.~J.,  {Lada} E.~A.,  2003, \mn@doi [\araa]
  {10.1146/annurev.astro.41.011802.094844}, \href
  {https://ui.adsabs.harvard.edu/abs/2003ARA&A..41...57L} {41, 57}

\bibitem[\protect\citeauthoryear{{Lucas} et~al.,}{{Lucas}
  et~al.}{2004}]{2004MNRAS.352.1347L}
{Lucas} P.~W.,  et~al., 2004, \mn@doi [\mnras]
  {10.1111/j.1365-2966.2004.08026.x}, \href
  {https://ui.adsabs.harvard.edu/abs/2004MNRAS.352.1347L} {352, 1347}

\bibitem[\protect\citeauthoryear{{Lucas}, {Hough}, {Bailey}, {Chrysostomou},
  {Gledhill}  \& {McCall}}{{Lucas} et~al.}{2005}]{2005OLEB...35...29L}
{Lucas} P.~W.,  {Hough} J.~H.,  {Bailey} J.,  {Chrysostomou} A.,  {Gledhill}
  T.~M.,   {McCall} A.,  2005, \mn@doi [Origins of Life and Evolution of the
  Biosphere] {10.1007/s11084-005-7770-6}, \href
  {https://ui.adsabs.harvard.edu/abs/2005OLEB...35...29L} {35, 29}

\bibitem[\protect\citeauthoryear{{Mathis}, {Rumpl}  \& {Nordsieck}}{{Mathis}
  et~al.}{1977}]{1977ApJ...217..425M}
{Mathis} J.~S.,  {Rumpl} W.,   {Nordsieck} K.~H.,  1977, \mn@doi [\apj]
  {10.1086/155591}, \href
  {https://ui.adsabs.harvard.edu/abs/1977ApJ...217..425M} {217, 425}

\bibitem[\protect\citeauthoryear{{McGuire}, {Carroll}, {Loomis}, {Finneran},
  {Jewell}, {Remijan}  \& {Blake}}{{McGuire}
  et~al.}{2016}]{2016Sci...352.1449M}
{McGuire} B.~A.,  {Carroll} P.~B.,  {Loomis} R.~A.,  {Finneran} I.~A.,
  {Jewell} P.~R.,  {Remijan} A.~J.,   {Blake} G.~A.,  2016, \mn@doi [Science]
  {10.1126/science.aae0328}, \href
  {https://ui.adsabs.harvard.edu/abs/2016Sci...352.1449M} {352, 1449}

\bibitem[\protect\citeauthoryear{{McKee} \& {Ostriker}}{{McKee} \&
  {Ostriker}}{2007}]{2007ARA&A..45..565M}
{McKee} C.~F.,  {Ostriker} E.~C.,  2007, \mn@doi [\araa]
  {10.1146/annurev.astro.45.051806.110602}, \href
  {https://ui.adsabs.harvard.edu/abs/2007ARA&A..45..565M} {45, 565}

\bibitem[\protect\citeauthoryear{Nakamura et~al.,}{Nakamura
  et~al.}{2022}]{nakamura2022origin}
Nakamura E.,  et~al., 2022, Proceedings of the Japan Academy, Series B, 98, 227

\bibitem[\protect\citeauthoryear{{Ouchi}, {Ono}  \& {Shibuya}}{{Ouchi}
  et~al.}{2020}]{2020ARA&A..58..617O}
{Ouchi} M.,  {Ono} Y.,   {Shibuya} T.,  2020, \mn@doi [\araa]
  {10.1146/annurev-astro-032620-021859}, \href
  {https://ui.adsabs.harvard.edu/abs/2020ARA&A..58..617O} {58, 617}

\bibitem[\protect\citeauthoryear{{Pizzarello} \& {Yarnes}}{{Pizzarello} \&
  {Yarnes}}{2018}]{2018P&SS..164..127P}
{Pizzarello} S.,  {Yarnes} C.~T.,  2018, \mn@doi [\planss]
  {10.1016/j.pss.2018.07.002}, \href
  {https://ui.adsabs.harvard.edu/abs/2018P&SS..164..127P} {164, 127}

\bibitem[\protect\citeauthoryear{{Roman-Duval}, {Jackson}, {Heyer}, {Rathborne}
   \& {Simon}}{{Roman-Duval} et~al.}{2010}]{2010ApJ...723..492R}
{Roman-Duval} J.,  {Jackson} J.~M.,  {Heyer} M.,  {Rathborne} J.,   {Simon} R.,
   2010, \mn@doi [\apj] {10.1088/0004-637X/723/1/492}, \href
  {https://ui.adsabs.harvard.edu/abs/2010ApJ...723..492R} {723, 492}

\bibitem[\protect\citeauthoryear{{Seon}, {Song}  \& {Chang}}{{Seon}
  et~al.}{2022}]{2022ApJS..259....3S}
{Seon} K.-i.,  {Song} H.,   {Chang} S.-J.,  2022, \mn@doi [\apjs]
  {10.3847/1538-4365/ac3af1}, \href
  {https://ui.adsabs.harvard.edu/abs/2022ApJS..259....3S} {259, 3}

\bibitem[\protect\citeauthoryear{Shoji, Kitazawa, Sato, Watanabe, Boero,
  Shigeta  \& Umemura}{Shoji et~al.}{2023}]{shoji2023enantiomeric}
Shoji M.,  Kitazawa Y.,  Sato A.,  Watanabe N.,  Boero M.,  Shigeta Y.,
  Umemura M.,  2023, The Journal of Physical Chemistry Letters, 14, 3243

\bibitem[\protect\citeauthoryear{{Tanaka}, {Tan}  \& {Zhang}}{{Tanaka}
  et~al.}{2016}]{2016ApJ...818...52T}
{Tanaka} K. E.~I.,  {Tan} J.~C.,   {Zhang} Y.,  2016, \mn@doi [\apj]
  {10.3847/0004-637X/818/1/52}, \href
  {https://ui.adsabs.harvard.edu/abs/2016ApJ...818...52T} {818, 52}

\bibitem[\protect\citeauthoryear{{Tazaki}, {Lazarian}  \& {Nomura}}{{Tazaki}
  et~al.}{2017}]{2017ApJ...839...56T}
{Tazaki} R.,  {Lazarian} A.,   {Nomura} H.,  2017, \mn@doi [\apj]
  {10.3847/1538-4357/839/1/56}, \href
  {https://ui.adsabs.harvard.edu/abs/2017ApJ...839...56T} {839, 56}

\bibitem[\protect\citeauthoryear{{Whitney} \& {Wolff}}{{Whitney} \&
  {Wolff}}{2002}]{2002ApJ...574..205W}
{Whitney} B.~A.,  {Wolff} M.~J.,  2002, \mn@doi [\apj] {10.1086/340901}, \href
  {https://ui.adsabs.harvard.edu/abs/2002ApJ...574..205W} {574, 205}

\bibitem[\protect\citeauthoryear{{Wolf}, {Voshchinnikov}  \& {Henning}}{{Wolf}
  et~al.}{2002}]{2002A&A...385..365W}
{Wolf} S.,  {Voshchinnikov} N.~V.,   {Henning} T.,  2002, \mn@doi [\aap]
  {10.1051/0004-6361:20020158}, \href
  {https://ui.adsabs.harvard.edu/abs/2002A&A...385..365W} {385, 365}

\bibitem[\protect\citeauthoryear{{Yajima}, {Umemura}, {Mori}  \&
  {Nakamoto}}{{Yajima} et~al.}{2009}]{2009MNRAS.398..715Y}
{Yajima} H.,  {Umemura} M.,  {Mori} M.,   {Nakamoto} T.,  2009, \mn@doi
  [\mnras] {10.1111/j.1365-2966.2009.15195.x}, \href
  {https://ui.adsabs.harvard.edu/abs/2009MNRAS.398..715Y} {398, 715}

\bibitem[\protect\citeauthoryear{{Yajima}, {Li}, {Zhu}  \& {Abel}}{{Yajima}
  et~al.}{2012}]{2012MNRAS.424..884Y}
{Yajima} H.,  {Li} Y.,  {Zhu} Q.,   {Abel} T.,  2012, \mn@doi [\mnras]
  {10.1111/j.1365-2966.2012.21228.x}, \href
  {https://ui.adsabs.harvard.edu/abs/2012MNRAS.424..884Y} {424, 884}

\bibitem[\protect\citeauthoryear{{Yajima}, {Li}, {Zhu}, {Abel}, {Gronwall}  \&
  {Ciardullo}}{{Yajima} et~al.}{2014}]{2014MNRAS.440..776Y}
{Yajima} H.,  {Li} Y.,  {Zhu} Q.,  {Abel} T.,  {Gronwall} C.,   {Ciardullo} R.,
   2014, \mn@doi [\mnras] {10.1093/mnras/stu299}, \href
  {https://ui.adsabs.harvard.edu/abs/2014MNRAS.440..776Y} {440, 776}

\bibitem[\protect\citeauthoryear{Yokoyama et~al.,}{Yokoyama
  et~al.}{2022}]{yokoyama2022samples}
Yokoyama T.,  et~al., 2022, Science, p. eabn7850

\makeatother
\end{thebibliography}
\input{main.bbl}




\appendix

\section{Calculation Method of Scattering} \label{appendix_scattering}

\begin{figure*}
	\begin{center}
	\includegraphics[width=140mm]{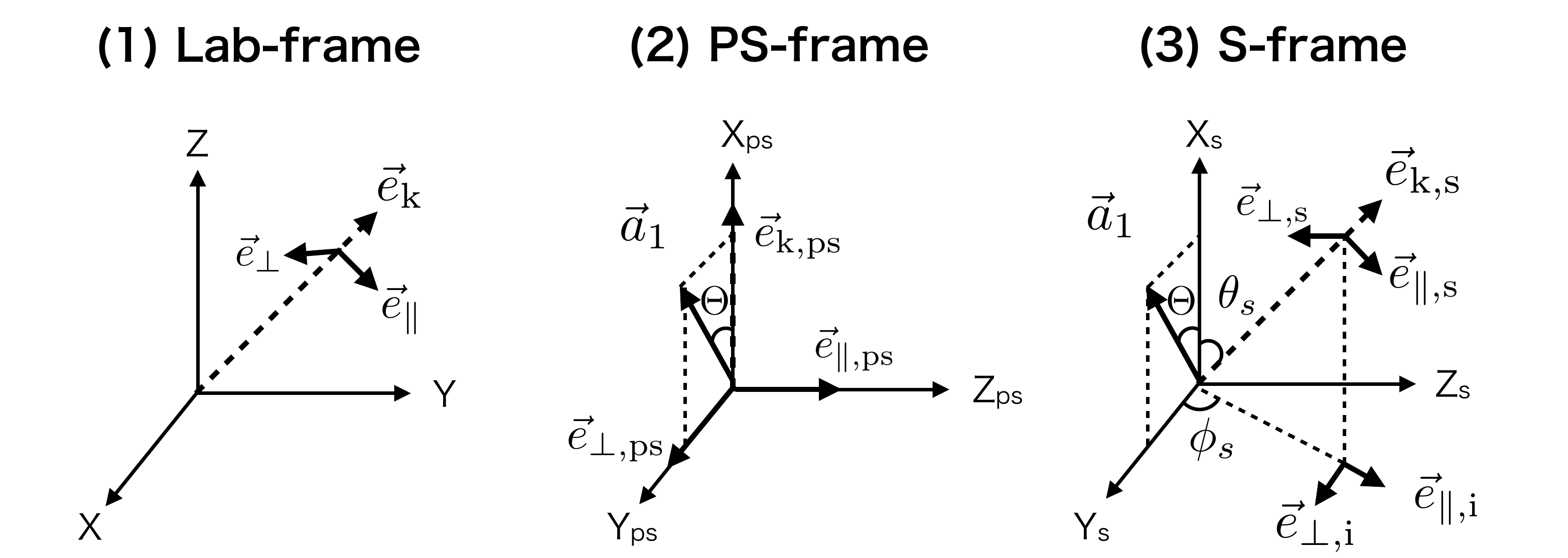}
	\end{center}
	\caption{
	The coordinates in the radiation transfer simulations; (1) the laboratory frame (Lab-frame), (2) the pre-scattering frame (PS-frame) and (3) the post-scattering frame (S-frame). 
	The vectors $\vec{e}_{\rm k}$, $\vec{e}_{\perp}$, and $\vec{e}_{\parallel}$ represent the directions of the photon packet, the parallel and perpendicular vectors defined in each coordinate.
	In the PS and S-frames, the vector $\vec{a}_1$ represents the direction of the minor axis of the dust grains, and $\Theta$ is the angle between the $\rm X_{\rm ps}$ and $\rm X_{\rm s}$ axes.
    $\theta_{\rm s}$ and $\phi_s$ represent scattering angles.
	}
	\label{zu1}
	\end{figure*}

In \citetalias{2020MNRAS.496.2762F}, we briefly summarized the calculation methods of scattering.
We describe the details of them in this Appendix.

The Stokes parameters $Q$ and $U$ depend on the orthonormal basis vectors.
In Equation \eqref{1.1}, the basis vectors are defined as the parallel basis vectors ($\vec{e}_{\parallel, i}$ and $\vec{e}_{\parallel, s}$) of the incident, and scattered light are contained in the same plane as shown in Fig \ref{zu1}-(3).
Here, we consider the three coordinates to simplify the coding as shown in Fig. \ref{zu1}:
\begin{description}
	\item[(1)] Lab-frame: We calculate the transportation of photon packets in the laboratory frame (Lab-frame). The parallel and perpendicular vectors of the electric field are defined as they coincide with the $\vec{e}_{\perp} = -\vec{e}_{\phi}$ and $\vec{e}_{\parallel} = \vec{e}_{\theta}$ where $\vec{e}_{\theta}$ and $\vec{e}_{\phi}$ are the unit vectors of the polar and azimuthal directions.
	\item[(2)] PS-frame: We use the PS-frame (pre-scattering frame) to calculate the scattering cross-section and determine the scattering direction. In this frame, the direction of the incident light is the same as the $\rm X_{\rm ps}$-axis. We set the direction of the minor axis of an oblate dust grain ($\vec{a}_{1}$) in the $\rm X_{\rm ps}$-$\rm Y_{\rm ps}$ plane at $z=0$. The $\rm Y_{\rm ps}$ and $\rm Z_{\rm ps}$ axes are parallel to the unit vectors of the incident light ($\vec{e}_{\perp}$ and $\vec{e}_{\parallel}$), respectively. 
	\item[(3)] S-frame: In the S-frame, we transform the Stokes parameters with the M\"uller matrix $\bf{M}$ defined as Equation \eqref{1.1}.  
    In this frame, the polar and azimuthal angles ($\theta_{\rm s}$ and $\phi_{\rm s}$) of the scattered light correspond to the scattering angles.
    The parallel vectors of the incident and scattering light are defined as coplanar. 
    The direction of the incident light is the same as the $\rm X_{\rm s}$-axis in this frame. 
\end{description}

In each cell, we calculate the cross-section of scattering $C_{\rm sca}$, which depends on the directions of the minor axis of a dust grain, and the incident and scattered light.
First, we alter the Stokes vector from Lab-frame $(\bf{I}_{\rm lab})$ to PS-frame $(\bf{I}_{\rm ps})$. 
The conversion of the Stokes vector is defined as \citep{1983asls.book.....B} 
\begin{eqnarray}
	\left( \begin{array}{c} 
	I^{'}\\
	Q^{'} \\ 
	U^{'} \\
	V^{'}
	\end{array} \right) =   \left(
	\begin{matrix} 
	1 & 0 & 0 & 0\\
	0 & \cos 2 \psi & \sin 2 \psi & 0  \\
	0 & - \sin  2 \psi & \cos 2 \psi & 0  \\
	0 & 0 & 0 & 1\\
	\end{matrix}
	\right)
	\left( \begin{array}{c} 
	I \\
	Q \\ 
	U \\
	V
	\end{array} \right), \label{A1.1}
\end{eqnarray}
where the angle $\psi$ is the clockwise rotation angle.
The rotation axis between the Lab and PS-frames ($\psi_{\rm lab \rightarrow ps}$) is obtained as follows.
In the PS-frame, the minor axis of a dust grain $\vec{a}_{1}$ is coplanar with the ${\rm X}_{\rm ps}$-${\rm Y}_{\rm ps}$ plane.
The projection vector of $\vec{a}_{1}$ to the ${\rm Y}_{\rm ps}$-${\rm Z}_{\rm ps}$ plane coincides with the perpendicular axis $\vec{e}_{\perp, \rm ps}$.
The projection unit vector of $\vec{a}_1$ to the ${\rm Y}_{\rm ps}$-${\rm Z}_{\rm ps}$ plane in the Lab-frame is obtained as
\begin{eqnarray}
	\vec{b} &=& \vec{a}_{1} - \left( \vec{a}_1 \cdot \vec{e}_{\rm k} \right) \vec{e_{\rm k}} \nonumber \\
		&=& \vec{a}_{1} - \cos \Theta \vec{e}_{\rm k} \label{A1.2}
\end{eqnarray}
where $\Theta$ is the angle between the incident light and the minor axis of a dust grain.
The perpendicular axis of the PS-frame in the Lab-frame is given as $\vec{e}_{\perp, {\rm ps}} = \vec{b}/|\vec{b}|$.
The rotation angle is obtained as
\begin{eqnarray}
	\psi_{\rm lab \rightarrow ps} = \begin{cases}
	\arccos(\vec{e}_{\rm \perp, ps} \cdot \vec{e}_{\rm \perp}) & (\vec{e}_{\rm \perp, ps} \cdot \vec{e}_{\rm \parallel} <  0) \\
	2 \pi - \arccos(\vec{e}_{\rm \perp, ps} \cdot \vec{e}_{\rm \perp}) & (\vec{e}_{\rm \perp, ps} \cdot \vec{e}_{\rm \parallel} \geqq  0).
	\end{cases}
	 \label{A1.3}
\end{eqnarray}

We use the M\"uller matrix $\bf{M}$ to determine the scattering direction in the PS-frame.
Since the matrix is originally defined in the S-frame, we derive the modified M\"uller matrix as
\begin{eqnarray}
	\bf{M}^{'} = \bf{M}
	\left(
	\begin{matrix} 
	1 & 0 & 0 & 0\\
	0 & - \cos 2 \phi_{\rm s} & \sin 2 \phi_{\rm s}   & 0  \\
	0 & - \sin 2 \phi_{\rm s} &  -\cos 2 \phi_{\rm s}& 0  \\
	0 & 0 & 0 & 1\\
	\end{matrix}
	\right),
	\label{A1.4}
\end{eqnarray}
where we adapt the rotational axis from the PS-frame to the S-frame as $\psi_{\rm ps \rightarrow s} = \pi /2 - \phi_{\rm s}$.
The differential cross-section of scattering is estimated with the modified M\"uller matrix $\bf{M}^{'}$ as
\begin{eqnarray}
	\frac{dC_{\rm sca}}{d \Omega} = \frac{1}{k^2} \frac{ ( S_{11}^{'} I_{\rm ps} + S_{12}^{'} Q_{\rm ps} + S_{13}^{'} U_{\rm ps} + S_{14}^{'} V_{\rm ps}) }{I_{\rm ps}}, \label{1.2}
\end{eqnarray}
where ${\bf I}_{\rm ps} =(I_{\rm ps}, Q_{\rm ps}, U_{\rm ps}, V_{\rm ps})$ is the Stokes vector in the PS-frame.
The scattering cross-section is given as
\begin{align}
	C_{\rm sca} = \frac{1}{k^2} \int d \Omega \left(\frac{S_{11}^{'} I_{\rm ps} + S_{12}^{'} Q_{\rm ps} + S_{13}^{'} U_{\rm ps} + S_{14}^{'} V_{\rm ps}}{I_{\rm ps}} \right). \label{cross_section_sca}
\end{align}
The scattering angles are randomly selected based on the probability from the differential cross-section in the PS-frame. Then, we convert the Stokes vector in the PS-frame ($\bf{I}_{\rm ps}$) to that in the S-frame ($\bf{I}_{\rm s}$) with the modified M\"uller matrix $\bf{M}^{'}$.
In our radiative transfer calculations, the intensity of a photon packet does not change in a scattering process. We change ratios of $Q$, $U$, and $V$ to $I$.
Finally, we convert the stokes vector $\bf{I}_{\rm s}$ to that in the Lab-frame $\bf{I}$ as in Equation \eqref{A1.3}.

\section{The observational map at the NIR wavelength}\label{sec:large_dust_grain_NIR}

\begin{figure*}
	\begin{center}
	\includegraphics[width=160mm]{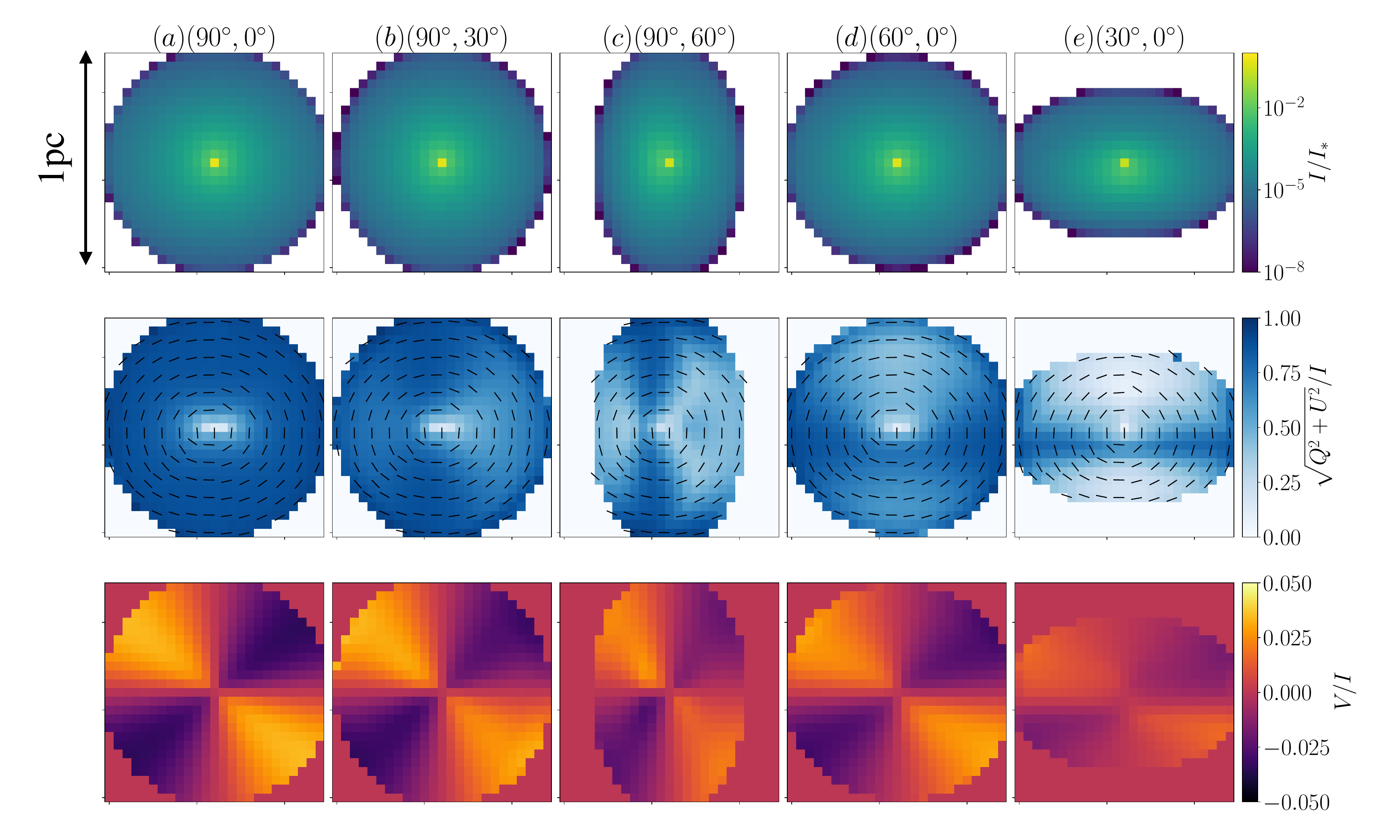}
	\end{center}
	\caption{
	Same as Fig. \ref{fig3}, but for the cases at $\lambda = 2.14~{\mu \rm m}$.
	}
	\label{fig2}
\end{figure*}

We present the results at the NIR wavelength of $\lambda = 2.14~{\rm \mu m}$.
In \citetalias{2020MNRAS.496.2762F}, we only discussed the face-on view.
Here, we simulate the observational maps of the viewing angles of $(\theta_{\rm obs}, \phi_{\rm obs})=(90^{\circ}, 0^{\circ})$, $(90^{\circ}, 30^{\circ})$, , $(90^{\circ}, 60^{\circ})$$(60^{\circ}, 0^{\circ})$, and $(30^{\circ}, 0^{\circ})$.
Fig. \ref{fig2} shows the observational maps from each angle.
Similar to the case of Ly$\alpha$, the intensity decreases outward.

The middle panels in Fig. \ref{fig2} show the linear polarization.
In the same fashion as seen in Fig. \ref{fig3}, the polarization vector shows the concentric pattern.
The degree of linear polarization mainly depends on the incident and scattered light angles represented in Fig. \ref{zu1} as $\theta_{\rm s}$.
At $(\theta_{\rm obs}, \phi_{\rm obs}) = (90^{\circ}, 0^{\circ})$, most of observed photons are perpendicularly scattered, which induces the highly linear polarization.
Thus, the degree of linear polarization reaches $\sim 1$ in the entire region.
On the other angles (b)-(e), the degree of linear polarization decreases in the horizontal or vertical directions of the slabs because the forward or backward photons are dominating.

The bottom panels of Fig. \ref{fig2} show the distributions of the CP degree at the NIR wavelength.
The symmetric quadrupole patterns appear.
These patterns are related to the dependence of the M\"uller matrix on the angles of $\Theta$, $\theta_{\rm s}$, and $\phi_{\rm s}$.
The value of $S_{41}/S_{11}$ represents the generation rates of the circular polarization from the non-polarized light.
The Rayleigh approximation is 
valid only for dust grains of which the size is much smaller than the wavelength of radiation. It is $<0.1~{\mu \rm m}$ for the NIR wavelength (see also Fig. \ref{fig:cp_sca_214}). 
As discussed in Appendix \ref{apdB}, the $S_{41}$ is estimated with the Rayleigh approximation as \citep[see Eq. \ref{eq_S41},][]{2000MNRAS.314..123G}
\begin{align}
    S_{41} \propto -(\alpha_1 \alpha_3^* - \alpha_1^* \alpha_3) \sin 2 \Theta \sin \theta \sin \phi. \label{eq_S41propto}
\end{align}
These patterns are consistent with the observed angular distributions of the CP degrees \citep[e.g.,][]{2013ApJ...765L...6K, 2014ApJ...795L..16K, 2016AJ....152...67K}.

\begin{figure*}
	\begin{center}
	\includegraphics[width=160mm]{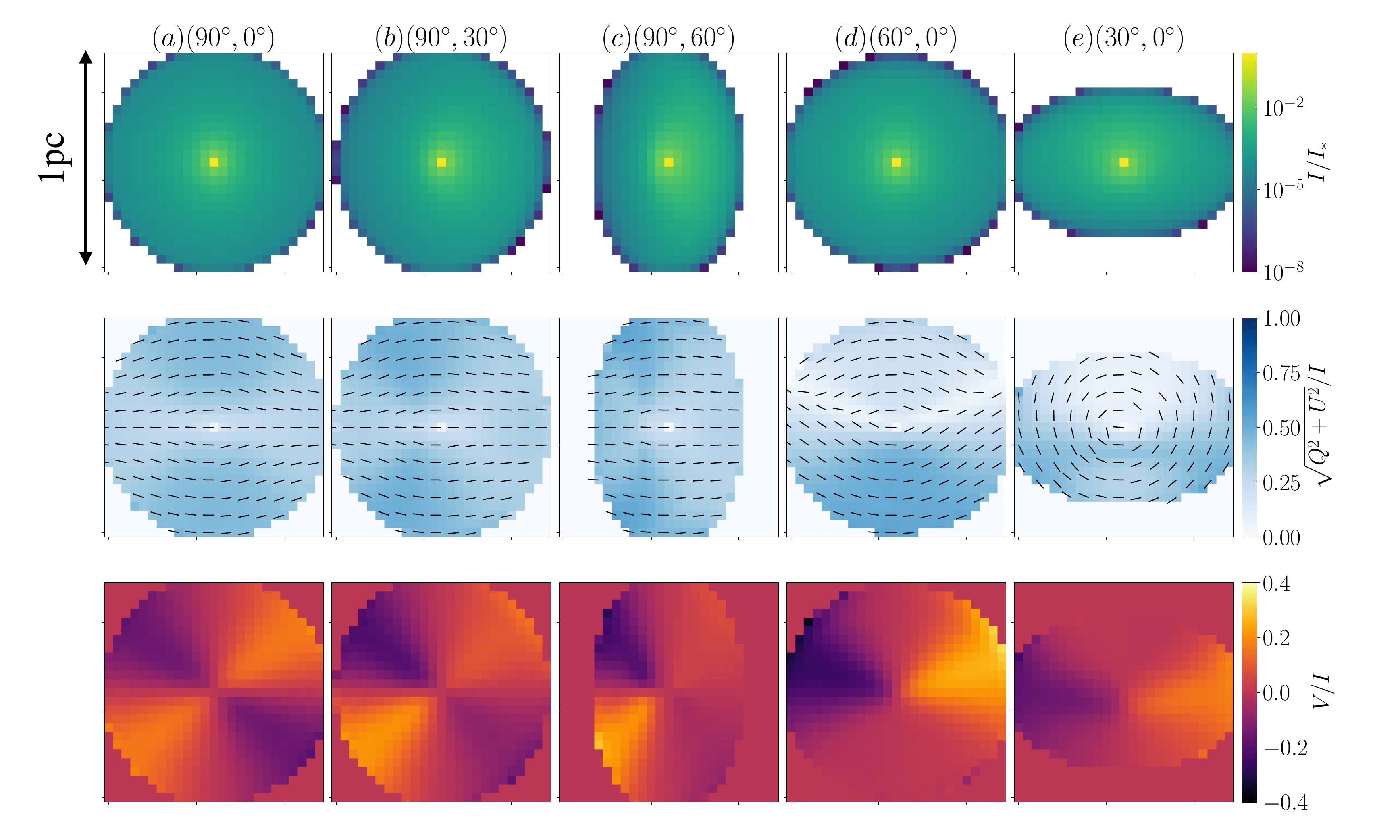}
	\end{center}
	\caption{
    Same as Fig. \ref{fig1}, but for the size distributions $n_{\rm d} \propto a_{\rm d}^{-3.5}$ and the size range $0.01~\mu {\rm m} \leqq a_{\rm d} \leqq 1~\mu {\rm m}$ at $\lambda = 2.14~{\mu \rm m}$. 
	}
	\label{fig1}
\end{figure*}

In \citetalias{2020MNRAS.496.2762F}, we showed that the CP degree exceeds 20 percent even at the NIR wavelength if the dust slab contains micron-size dust grains.
Here, we perform the additional simulations with the large dust grain size as $0.01~{\rm \mu m} \leqq a_{\rm d} \leqq 1~{\rm \mu m}$ and investigate the observational maps with the different viewing angles. 
Figure \ref{fig1} shows the intensity, linear, and circular polarization maps. 
In the face-on view, the CP degree exceeds 20 percent in the diagonal direction.
The sign of the degree is reversed compared with the fiducial model of Figure \ref{fig2}, which is caused by the micron-size dust grains as shown in \citetalias{2020MNRAS.496.2762F}.
The distributions of CP are the asymmetric quadrupole patterns in the edge-on view of oblate grains (Fig.\ref{fig1}-a, b, and c).
However, in the cases of $(\theta_{\rm obs}, \phi_{\rm obs})=(60^\circ, 0^\circ)$ and $(30^\circ, 0^\circ)$, the dipole patterns of CP light appears unlike the fiducial model. 
Thus, the CP pattern of the NIR wavelength is not universal if the micron-size dust grains contained.
The dipole pattern can be a probe of the dust size extending $\sim 1~\rm \mu m$.

\section{Rayleigh approximation}\label{apdB}

Here, we derive the M\"uller matrix for an oblate dust grain based on the Rayleigh approximation.
The scattering matrix was calculated in Appendix A of \citet{2000MNRAS.314..123G}, but the coordinates adapted in their formula are different from in Fig. \ref{zu1}.
Here, we reconstruct the formula to clarify.

The incident electric field causes the dipole moment as
\begin{align}
    \bf{p} = \alpha \bf{E}, \label{eq_dipole}
\end{align}
where $\alpha$ is the polarizability tensor.
If the oblate dust grain axis coincides with the coordinate axis, $\alpha$ only has the diagonal components.
Each diagonal represents as \citep{1983asls.book.....B} 
\begin{align}
    \alpha_{i} = 4 \pi a_1^2 a_3 \frac{(m^2-1)}{3+3L_{i}(m^2-1)}, \label{eq_alpha}
\end{align}
where $m$, $a_1$, and $a_3$ are the refractive index, the radii of the major and minor axes.
The shape factors $L_{i}$ of oblate satisfy $L_1=L_2<L_3$ and $L_1 + L_2 + L_3 = 1$.
The shape factor in the direction of the major axis $L_1$ is given as  \citep{1983asls.book.....B}
\begin{align}
    L_1 = \frac{g(e)}{2 e^2} \left[ \frac{\pi}{2} - \arctan(g(e)) \right] - \frac{g(e)^2}{2}, \label{eq_L1}
\end{align}
where
\begin{align}
    g(e) = \left( \frac{1-e^2}{e^2} \right)^{1/2}, \label{eq_ge}
\end{align}
and
\begin{align}
    e^2 = 1 - a_1^2/a_3^2. \label{eq_e2}
\end{align}

The scattered light is obtained as 
\begin{align}
    {\bf E}_{\rm s} = \frac{k^2 e^{ikr}}{4 \pi r} ({\bf e}_{{k,s}} \times {\bf p}) \times  {\bf e}_{{k,s}}, 
\end{align}
where $k$, $r$ and ${\bf e}_{\bf r}$ are the photon wavenumber, the distance from the scattered point, and the direction of the scattered light.
Here, we adapt the coordinate defined as Fig. \ref{zu1}.
The basis vectors of incident light $({\bf{e}}_{k,i}, {\bf e}_{\perp,i}, {\bf e}_{\parallel,i})$ and scatted light $({\bf{e}}_{k,s}, {\bf e}_{\perp, s}, {\bf e}_{\parallel, s})$ are given as
\begin{align}
{\bf e}_{k,i} = 
\left( \begin{array}{c} 
1 \\
0 \\
0
\end{array} \right), 
{\bf e}_{\perp,i} = 
\left( \begin{array}{c} 
0 \\
\sin \phi_{\rm s} \\
-\cos \phi_{\rm s}
\end{array} \right), 
{\bf e}_{\parallel,i} = 
\left( \begin{array}{c} 
0 \\
\cos \phi_{\rm s} \\
\sin \phi_{\rm s}
\end{array} \right), 
\end{align}
and
\begin{align}
{\bf e}_{k,s} = 
\left( \begin{array}{c} 
\cos \theta_{\rm s} \\
\sin \theta_{\rm s} \cos \phi_{\rm s} \\
\sin \theta_{\rm s} \sin \phi_{\rm s}
\end{array} \right), 
{\bf e}_{\perp,s} = 
\left( \begin{array}{c} 
0 \\
\sin \phi_{\rm s} \\
- \cos \phi_{\rm s}
\end{array} \right),
{\bf e}_{\parallel,s} = 
\left( \begin{array}{c} 
- \sin \theta_{\rm s} \\
\cos \theta_{\rm s} \cos \phi_{\rm s} \\
\cos \theta_{\rm s} \sin \phi_{\rm s} 
\end{array} \right). 
\end{align}
Then, the scattered light is calculated as 
\begin{align}
    {\bf E}_{\rm s} &= F \left ( p_{\rm Y} \sin \phi_{\rm s} - p_{\rm Z} \cos \phi_{\rm s}  \right) {\bf e}_{\perp, s}   \nonumber \\
    & + F \left( - p_{\rm X} \sin \theta_{\rm s}  + p_{\rm Y} \cos \theta_{\rm s} \cos \phi_{\rm s}  +  p_{\rm Z} \cos \theta_{\rm s} \sin \phi_{\rm s} \right) {\bf e}_{\parallel, s} \label{eq_Es}
\end{align}
where 
\begin{align}
    F = \frac{k^2 e^{ikr}}{4 \pi r}. \label{eq_F}
\end{align}
The dipole moment ${\bf p} = (p_{\rm X}, p_{\rm Y}, p_{\rm Z})$ is obtained by transforming the diagonal matrix ${\boldsymbol \alpha}$ into the coordinate system of (X, Y, Z).
The transforming matrix corresponds to rotation by an $\Theta$ angle around the axis Z.
The dipole moment is given as
\begin{align}
 {\bf p} = R_z (\Theta) \mathcal{A} R_z(-\Theta) {\bf E}_{i} 
 = \left( 
\begin{array}{c}
P_{11} E_{\perp, i} + P_{12} E_{\parallel, i} \\
P_{21} E_{\perp, i} + P_{22} E_{\parallel, i} \\
P_{31} E_{\perp, i} + P_{32} E_{\parallel, i} \\
\end{array}
 \right), \label{eq:Eq_p_xyz} 
\end{align}
where $R_z (\Theta)$ and $\mathcal{A}$ are defined as
\begin{align}
    \mathcal{A} = \begin{pmatrix} \alpha_{3} & 0 & 0 \\ 0 & \alpha_{1} & 0 \\ 0 & 0 & \alpha_{1} \end{pmatrix}, \label{eq:definition_of_mathcal_A}
\end{align}
and
\begin{align}
R_z (\Theta) = 
\left( \begin{matrix} 
\cos \Theta & - \sin \Theta & 0 \\
\sin \Theta & \cos \Theta & 0 \\
0 & 0 & 1
\end{matrix} \right). \label{eq:definition_of_Rz}
\end{align}
The incident electric field is estimated with the basis vectors $( {\bf e}_{\perp,i}, {\bf e}_{\parallel,i})$ as
\begin{align}
    {\bf E}_{i} = E_{\perp, i}{\bf e}_{\perp, i}+E_{\parallel, i}{\bf e}_{\parallel, i}= \begin{pmatrix} 0 \\ E_{\perp, i} \sin \phi_{\rm s} + E_{\parallel, i} \cos \phi_{\rm s} \\ - E_{\perp, i} \cos \phi_{\rm s} + E_{\parallel, i} \sin \phi_{\rm s} \end{pmatrix}. \label{eq:incident_elec_field}
\end{align}
Substituting equations \eqref{eq:definition_of_mathcal_A}, \eqref{eq:definition_of_Rz}, and \eqref{eq:incident_elec_field} into \eqref{eq:Eq_p_xyz}, we obtain each factor for the dipole moment $\bf p$ as
\begin{align}
    P_{11} &= a \sin \phi_{\rm s}, \label{eq:P11} \\ 
    P_{12} &= a \cos \phi_{\rm s}, \label{eq:P_12}\\ 
    P_{21} &= b \sin \phi_{\rm s}, \label{eq:P_21}\\
    P_{22} &= b \cos \phi_{\rm s}, \label{eq:P_22}\\ 
    P_{31} &= -c \cos \phi_{\rm s}, \label{eq:P_31}\\ 
    P_{32} &= c \sin \phi_{\rm s},  \label{eq:P_32}
\end{align}
where
\begin{align}
    a &= (\alpha_3 - \alpha_1) \cos \Theta \sin \Theta, \label{eq:def_a} \\
    b &= \alpha_1 \cos^2 \Theta + \alpha_3 \sin^2 \Theta,  \label{eq:def_b} \\ 
    c &= \alpha_{1}. \label{eq:def_c}
\end{align}

The relation between the incident and scattered light is written with the amplitude scattering matrix as \citep{1983asls.book.....B}, 
\begin{align}
    \left( \begin{array}{c}
    E_{\parallel} \\
    E_{\perp}
    \end{array} \right)_{s} = \frac{e^{ik(r-x)}}{-ikr} 
    \left( \begin{matrix}
    S_2 & S_3 \\
    S_4 & S_1
    \end{matrix}
    \right)   \left( \begin{array}{c}
    E_{\parallel} \\
    E_{\perp}
    \end{array} \right)_{i}, \label{eq_amplitude_scat_matrix}
\end{align}
where we ignore the time-dependent term.
Substituting equations \eqref{eq_Es} and \eqref{eq:Eq_p_xyz} into \eqref{eq_amplitude_scat_matrix}, we obtain the each components of the amplitude scattering matrix as 
\begin{align}
    S_1 &= \frac{k^3}{4 \pi i} \left( b \sin^2 \phi_{\rm s} + c \cos^2 \phi_{\rm s}  \right),  \label{eq_S1} \\
    S_2 &= \frac{k^3}{4 \pi i} \left( -a \sin \theta_{\rm s} \cos \phi_{\rm s} + b \cos \theta_{\rm s} \cos^2 \phi_{\rm s} + c \cos \theta_{\rm s} \sin^2 \phi_{\rm s} \right) , \label{eq_S2} \\
    S_3 &= \frac{k^3}{4 \pi i} \left(-a \sin \theta_{\rm s} \sin \phi_{\rm s} + b \cos \theta_{\rm s} \cos \phi_{\rm s} \sin \phi_{\rm s} -c \cos \theta_{\rm s} \cos \phi_{\rm s} \sin \phi_{\rm s} \right), \label{eq_S3} \\
    S_4 &= \frac{k^3}{4 \pi i} \left(  b \cos \phi_{\rm s} \sin \phi_{\rm s} - c \cos \phi_{\rm s} \sin \phi_{\rm s}  \right). \label{eq_S4}
\end{align}
The $S_{11}$ and $S_{41}$ components of M\"uller matrix defined as Equation \eqref{1.1} are calculated with the amplitude scattering matrix as
\begin{align}
    S_{11} &= \frac{1}{2} \left( |S_1|^2+|S_2|^2+|S_3|^3+|S_4|^4 \right) \nonumber \\
    & = \frac{k^6}{32 \pi^2} \left[ f_{11} |\alpha_1|^2 + f_{31} \left( \alpha_1 \alpha_3^* + \alpha_1^* \alpha_3 \right) + f_{33} |\alpha_3|^2 \right], \label{eq_S11}
\end{align}
where 
\begin{align}
    f_{11} &= \cos^2 \phi_{\rm s} + \cos^2 \theta_{\rm s} \sin^2 \phi_{\rm s} \nonumber \\
    & + \cos^2 \Theta [(\sin \theta_{\rm s} \sin \Theta + \cos \theta_{\rm s} \cos \phi_{\rm s} \cos \Theta )^2 + \sin^2 \phi_{\rm s} \cos ^2 \Theta], \label{eq:f11} \\
    f_{31} &= \cos \Theta \sin \Theta [(\cos^2 \theta_{\rm s} - \sin^2 \theta_{\rm s} \cos^2 \phi_{\rm s}) \cos \Theta \sin \Theta \nonumber \\ 
    &-\cos \theta_{\rm s} \sin \theta_{\rm s} \cos \phi_{\rm s} (\cos ^2 \Theta - \sin^2 \Theta)], \label{eq:f31} \\
    f_{33} &= \sin^2 \Theta [(\sin \theta_{\rm s} \cos \Theta - \cos \theta_{\rm s} \cos \phi_{\rm s} \sin \Theta )^2 + \sin^2 \phi_{\rm s} \sin ^2 \Theta], \label{eq:f33}
\end{align}
and 
\begin{align}
    S_{41} &= {\rm Im} \left( S_2^*S_4 + S_3^* S_1 \right) \nonumber \\
    &= -\frac{k^6}{64 \pi^2} {\rm Im}(\alpha_1 \alpha_3^* - \alpha_1^* \alpha_3) \sin \theta_{\rm s} \sin \phi_{\rm s} \sin 2 \Theta. \label{eq_S41}
\end{align}

\begin{figure}
	\begin{center}
	\includegraphics[width=\columnwidth]{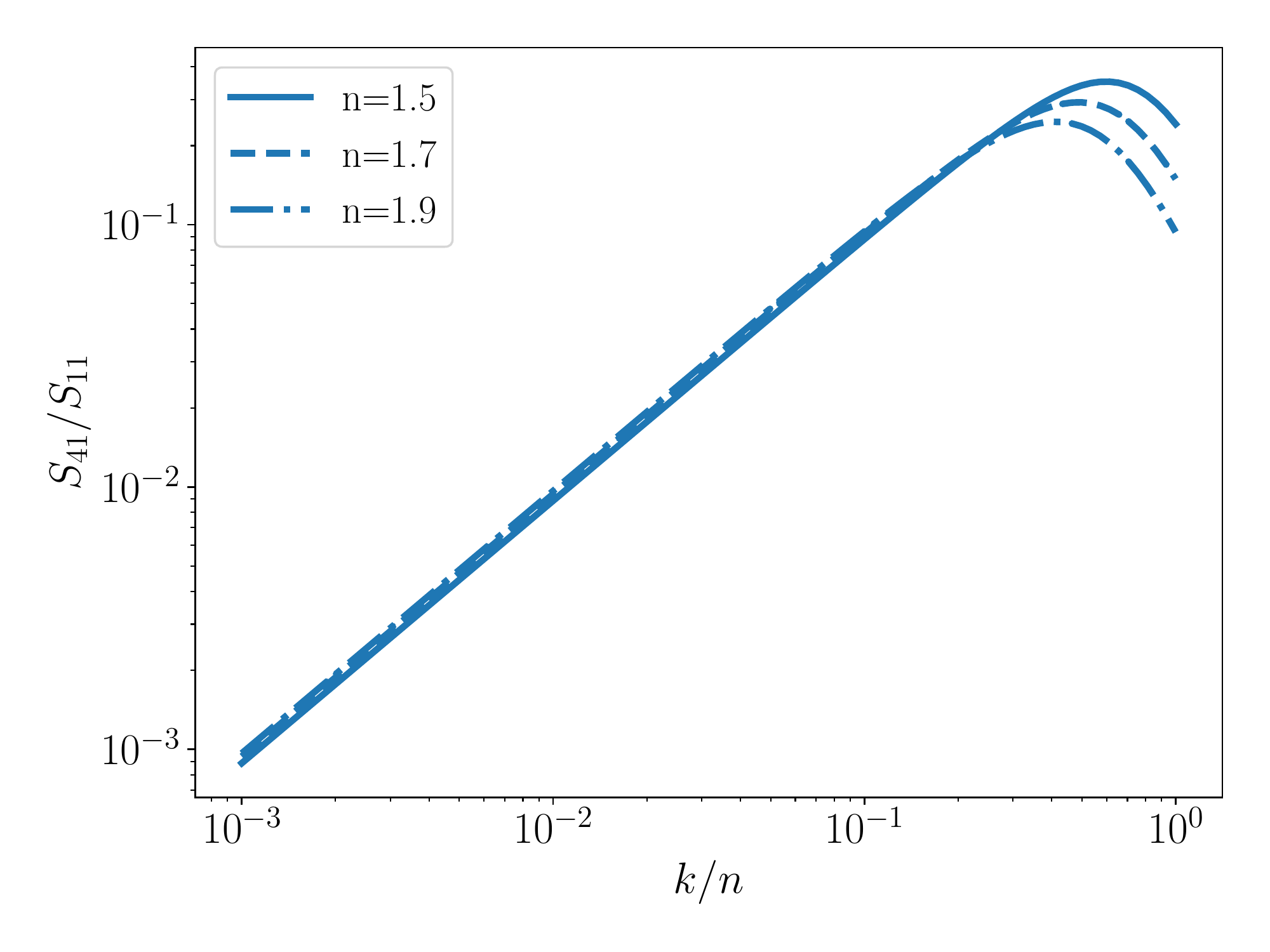}
	\end{center}
	\caption{
	The relation between $S_{41}/S_{11}$ and the imaginary part of the refractive index $k$.
	Each line represents cases with $n=1.5$ (solid), $1.7$ (dashed), and $1.9$ (dot-dashed).
	}
	\label{fig_S11S41_nk}
\end{figure}

Fig. \ref{fig_S11S41_nk} shows $S_{41}/S_{11}$ with the various  refractive index $m=n+ik$ with the scattering angle and the inclined angle of the dust grain $(\theta_{\rm s}, \phi_{\rm s}) = (90^{\circ}, 270^{\circ})$ and $\Theta = 45^{\circ}$.
Here, we adopt the oblate dust grain with the axial ratio of 2:1. 
The value of $S_{41}/S_{11}$ only depends on the ratio of the imaginary part to the real part of the refractive index.
At $k/n < 0.5$, the value of $S_{41}/S_{11}$ increases with $k/s$.
The refractive index of astronomical silicate is $k/n = 0.02$ for $\lambda = 2.14~{\rm \mu m}$ and $k/n = 0.37$ for $\lambda = 0.1216~{\rm \mu m}$.
Thus, the CP degree of the scattered Ly$\alpha$ photons is more than 5 times higher than that of the NIR light on the Rayleigh approximation.

\bsp	
\label{lastpage}
\end{document}